\documentclass[journal=mamobx,manuscript=article]{achemso}
\usepackage{titlesec}
\usepackage{fancybox}
\usepackage{mathrsfs}

\usepackage{setspace}
\usepackage{here}

\usepackage{comment}
\usepackage{enumerate}
\usepackage[dvips]{color}
\graphicspath{{./eps/}}

\usepackage{amsmath}
\usepackage{cases}

\title{Modeling induction period of polymer crystallization}
\author{Hiroshi Yokota}
\email{yokota@cmpt.phys.tohoku.ac.jp}
\author{Toshihiro Kawakatsu}
\affiliation{Department of Physcs, Tohoku University, Sendai, Japan}

\begin{document}

\maketitle
\begin{figure}[H]
\begin{center}
\includegraphics[clip, width=11cm,height=4.9cm]{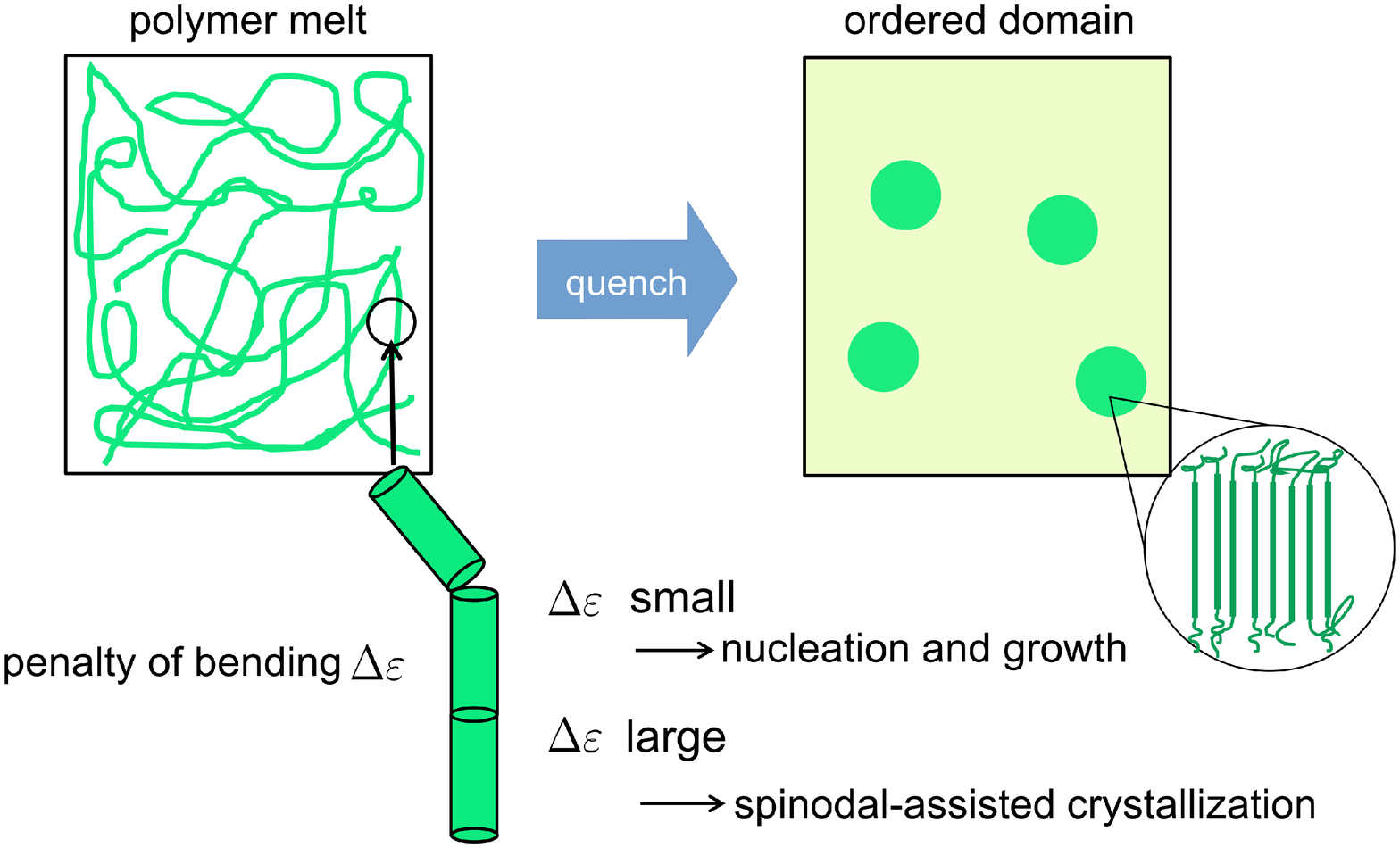}
\end{center}
\end{figure}
\newpage
\section*{Abstract}
We study the possibility of the spinodal decomposition in the induction period of the polymer crystallization. This phenomenon was first reported in an X-ray scattering experiment, and has still been controversial due to various experiments and theories that support or deny the phenomenon. 
In this article, we explain the condition for the spinodal decomposition to occur in polymer melts by deriving a Ginzburg-Landau model of the free energy as a functional of the density and the orientation of the segments, where we introduce the excluded volume and the nematic interactions through a combination of the random phase approximation and the transfer matrix for the polymer conformation. 
We show that, upon elimination of the degrees of freedom of the orientation, the nematic interaction reduces to an effective attraction whose strength increases with the stiffness of the polymer chain. Such an attraction induces spinodal decomposion especially for stiff polymer chain case.

\section{Introduction}\label{sec:introduction}
A polymer crystallization is a complex phenomenon where a positional order and an orientation order of segments play an important role. This is in contrast to the case of a simple liquid where the knowledge of only the positional order is required to understand the phenomenon\cite{Alder}.
In addition, the knowledge on the conformation entropy of the polymer chains and the interaction between segments\cite{Yamamoto2008} is also indispensable to the discussions on the polymer crystallization.
Originating from such a complexity, polymer crystals in general have a hierarchical structure over a wide range of length scales.  The structure on the smaller lengths scale (c.a. 10-100 nm) is the lamellar crystal structure formed by alternate layers of disordered chains and folded chains. Such a lamellar crystal grows radially, leading to the so-called ``spherulite'' structure ($\atop{>}{\sim} \mu {\rm m}$ ).

Although there have been researches that focus only on a single structure, for example the researches on the lamellar crystal structure\cite{milner} or those on the spherulite\cite{phase}, no framework that connects the structures on different length scales in the hierarchy has been reported.  In such a multiscale modelling, it is necessary to discuss the correlation between the microscopic structures and the mesoscopic (or macroscopic) structures which is expected to play an important role in the mechanism of the ordering processes of the crystallization.
Our final goal is to construct a model which can explain the mechanism of the whole structures in the hierarchy in a unified manner.  

As the first step to this goal, we focus on the phase separation process in the early stage of the polymer crystallization.
In general, the dynamics of the early stage of the phase separation is classified into two mechanisms, {\it i.e.} ``nucleation and growth (NG)'' and ``spinodal decomposition (SD)''.
Each of these two mechanisms occurs depending on the external condition such as the quench depth.
A shallow quench leads to the NG process as a result of 1st order phase transition, while a deep quench to the SD process as an example of the 2nd order phase transition.
When we change some of the control parameters to induce the phase transition from liquid to solid, it is believed that an NG occurs due to the different symmetries between the two phases.
In an NG process, small nuclei are initially formed by thermal fluctuation.
When the size of a nucleus becomes larger than a certain critical size where the bulk energy of the nucleus overcomes the surface energy, this nucleus grows.
Such a nucleus is called ``critical nucleus''.
The ``induction period'' is defined as the time regime before a critical nucleus is generated.

In the early 1990s, Imai {\it et al}. discovered interesting phenomena during the induction period of polymer crystallization by using X-ray scattering experiments on poly(ethylene terephthalate)(PET)\cite{imai1}\cite{imai2}\cite{imai3}.
According to their reports, amplitude of the long wavelength fluctuations of density (represented by a peak at $0.4$ nm$^{-1}$) increases exponentially with time, which implies that the SD occurs as a sign of the 2nd order phase transition.
It should be noted that such an SD-like behavior was observed in the induction period prior to the emergence of the Bragg peaks of the crystal structure. 
Imai {\it et al.} theoretically interpreted this phenomenon using a theory of liquid crystalline polymer where they included a coupling between the density and the orientation of the segment\cite{Doi1}\cite{Doi2}\cite{Doi3}.
The validity of their interpretation was confirmed by another experiment using depolarized light scattering(DPLS) technique\cite{Matsuda}, where they revealed that an orientation order grows during the induction period\cite{Matsuda}.
In response to these experimental data which imply SD, Gee {\it et al}. demonstrated, by using a large scale molecular dynamics(MD) simulation, that the SD occurs in the induction period, where the persistence length of the polymer chains increases\cite{Gee}. 
In analytical approaches Olmsted {\it et al}. proposed a phenomenological model  for the SD of the polymer crystallization focusing on the static properties of polymer melts\cite{Olmsted}.
In this work by Olmsted {\it et al.}, the authors chose the density fluctuation and the chain conformation as the order parameters and proposed a Ginzburg-Landau(GL) free energy model.
They discussed the chain conformation by using the local distribution of dihedral angles between consecutive covalent bonds of the backbone.
By assuming that an increase in the segment density prefers a nematic order, Olmsted {\it et al}. drew binodal and spinodal lines in the phase diagram, where they implicitly assumed the SD process in their model.
Extending their static model to dynamics, Tan {\it et al}. simulated the phase separation process based on the time dependent GL (TDGL) equation\cite{Tan}.

Contrary to the above experimental and theoretical evidences on the SD, there have been arguments that posed a question on these results.  For example, Wang {\it et al}. performed a small-angle X-ray scattering(SAXS) experiment and a light scattering experiment on poly(propylene)\cite{Z-G-Wang}, and compared their scattering data in the induction period with those of Cahn-Hilliard theory\cite{Cahn-Hilliard} and Avrami theory\cite{Avrami}. It is well-known that Cahn-Hilliard theory and Avrami theory can describe SD and NG processes, respectively. As a result, Wang {\it et al.} concluded that Avrami theory is appropriate to the polymer crystallization and denied that possibility of SD for this process. The SAXS data reported by Wang {\it et al.} showed an increase in its peak intensity but with constant peak position even in the late stage, which implies that SD does not occur in the induction period.
In order to explain the discrepancy between the experimental results by Imai {\it et al}. and their results, Wang {\it et al.} attributed the SD observed by Imai {\it et al.} to the low sensitivities of the X-ray scattering experiments to the density and orientation fluctuations. 
Panine {\it et al}. confirmed these results by Wang {\it et al.} using SAXS and wide-angle X-ray scattering(WAXS) techniques where they improved the accuracy of the X-ray scattering experiment compared with the previous ones\cite{Panine}. 
In these experiments, Panine {\it et al.} did not observe an SD in the induction period of the polymer crystallization, which agrees with the conclusion by Wang {\it et al.}.

In response to these controversial discussions on the mechanism of the phase separation in the induction period, Chuang {\it et al.} studied the detail of an ordering process in the early stage of the polymer crystallization by using several techniques such as SAXS, WAXS, Fourier transform infrared spectroscopy, and small-angle light scattering\cite{Chuang}.
Chuang {\it et al.} suggested that the mechanism of the phase separation depends on some parameters of the samples, {\it i.e.}, crystallization temperature, conformation of the chain in the initial condition and so on\cite{Chuang}.
  
Summarizing the above-mentioned researches on the behavior in the induction period, a deep quench (Imai {\it et al.}) seems to induce SD while a shallow quench like in the experiments by Wang {\it et al.} and by Panine {\it et al.} does not lead to SD.
Furthermore, we expect that the stiffness of the polymer chain is also an important factor for the SD as was proposed by Imai {\it et al.} \cite{imai1995} and by Gee {\it et al.}\cite{Gee}. Such an effect of the stiffness was not considered in the theoretical work by Olmsted {\it et al.}\cite{Olmsted}.

To construct a complete theoretical framework for the phenomena, we need to reconsider the condition of the SD without a priori assumptions used by Olmsted {\it et al.} and others. 
Our strategy is as follows.  First we derive the GL free energy by expanding the total free energy around the initial uniform state with respect to the density and the orientation fluctuations of the segments.
Analyzing the 2nd order terms in the GL expansion gives us the information on the stability of the initial uniform melt. 
It is commonly understood that expanding free energy with respect to only density fluctuation does not lead to the SD in a single component system.
To reproduce the SD, therefore, it is essential to introduce another degree of freedom coupled to the density fluctuation, such as orientation fluctuation in the GL expansion.
To calculate the expansion coefficients, which is described in terms of the density and the orientation correlation functions of the segments, we need to know the statistical properties of chain conformations.  Such information can be obtained by using the transfer matrix technique.

The present paper is organized as follows.
In the next section, we derive the GL free energy model for the instability of polymer melts, where we also discuss the effects of the coupling between the density and the orientation of the segments.
Using the free energy model, we reproduce the SD and estimate the correlation length of the density fluctuation during the induction period in Sec.\ref{sec:result}.
In the same section, we also interpret the correlation length based on the Cahn-Hilliard type linearlized theory on our model.
Finally, we conclude our results in Sec.\ref{sec:conclusion}.

\section{Model}\label{sec:model}
\subsection{Strategy of the model}
To study the polymer crystallization, we consider the situation where a single component polymer system is quenched from an initial high temperature melt state to a low temperature state below the crystallization temperature.

In order to study the dynamics of crystallization, we derive a free energy model of the system by expanding it with respect to the fluctuations of the segment density and the segment orientation around a uniformly mixed melt.
 
First, we consider the initial uniform and isotropic polymer melt state as a reference state of the expansion.
Because of the screening effect which originates from the incompressibility of the melt\cite{Kawakatsu}, the uniform polymer melt can be modelled as a set of independent ideal chains with an incompressibility condition. 
In the present study, we model the polymer chain as a sequence of rod-like segments in order to incorporate the nematic interaction.

In order to describe our model Hamiltonian for the uniform melt state, we introduce microscopic variables, {\it i.e.} the position $\mbox{\boldmath$r$}^{(i)}$ and the orientation vector $\mbox{\boldmath$b$}^{(i)}$ of $i$-th segment, which are denoted as $\Gamma \equiv \{ \{ \mbox{\boldmath$r$}^{(i)} \}, \{ \mbox{\boldmath$b$}^{(i)} \} \}$ in short.  Using these microscopic variables, the local density fluctuation $\delta\hat{\phi}(\mbox{\boldmath$r$};\Gamma)$ and the local orientation order parameter of the segments $\hat{S}_{\alpha\beta}(\mbox{\boldmath$r$};\Gamma)$, respectively, are defined as
\begin{align}
\delta\hat{\phi}(\mbox{\boldmath$r$})&=\sum_{i=0}^N \delta(\mbox{\boldmath$r$}-\mbox{\boldmath$r$}^{(i)})-\bar{\phi},\label{eqn:density}\\
\hat{S}_{\alpha\beta}(\mbox{\boldmath$r$})&=\frac{3}{2}\sum_{i=0}^N\left(b_{\alpha}^{(i)}b_{\beta}^{(i)}-\frac{1}{3} \right)\delta\left(\mbox{\boldmath$r$}-\mbox{\boldmath$r$}^{(i)} \right),\label{eqn:orientation}
\end{align}
where $b$ is the length of the rod-like segment and the symbol $\hat{\cdots}$ means that the variable $\cdots$ is a function of the phase space point $\Gamma$.

In terms of these coarse-grained field variables, Hamiltonian of the system can be written as:
\begin{align}
\hat{\mathcal{H}}&=\hat{\mathcal{H}}_0+\hat{\mathcal{W}}\nonumber\\
&=\hat{\mathcal{H}_0}+\hat{\mathcal{W}}_\phi[\hat{\phi}(\mbox{\boldmath$r$})]+\hat{\mathcal{W}}_S[\hat{S}_{\alpha\beta}(\mbox{\boldmath$r$})]\nonumber\\
&=\hat{\mathcal{H}}_0+\frac{v}{2}\int d\mbox{\boldmath$r$}\left\{ \delta \hat{\phi}(\mbox{\boldmath$r$};\Gamma)\right\}^2 +\frac{V_{\alpha\beta\alpha^\prime\beta^\prime}}{2}\int d\mbox{\boldmath$r$}\hat{S}_{\alpha\beta}(\mbox{\boldmath$r$};\Gamma)\hat{S}_{\alpha^\prime\beta^\prime}(\mbox{\boldmath$r$};\Gamma),\label{eqn:Hamiltonian}
\end{align}
where $\hat{\mathcal{H}}_0$ is the Hamiltonian for the reference uniform and isotropic melt, $\hat{\mathcal{W}}$ is the deviation from the reference state, and the functionals $\hat{\mathcal{W}}_S[\hat{S}_{\alpha\beta}(\mbox{\boldmath$r$})]$ and $\hat{\mathcal{W}}_\phi[\hat{\phi(\mbox{\boldmath$r$})}]$ are defined as
\begin{align}
\hat{\mathcal{W}}_S[\hat{S}_{\alpha\beta}(\mbox{\boldmath$r$})]&=\frac{V_{\alpha\beta\alpha^\prime\beta^\prime}}{2}\int d\mbox{\boldmath$r$} \hat{S}_{\alpha\beta}(\mbox{\boldmath$r$})\hat{S}_{\alpha^\prime\beta^\prime}(\mbox{\boldmath$r$})\\
\hat{\mathcal{W}}_\phi[\hat{\phi}(\mbox{\boldmath$r$})]&=\frac{v}{2}\int d\mbox{\boldmath$r$} \left\{\hat{\delta\phi}(\mbox{\boldmath$r$})  \right\}^2.
\end{align}
We expand $\hat{\mathcal{W}}$ in a power series in $\delta \hat{\phi}$ and $\hat{S}$ and retain up to the 2nd order terms.
The subscripts $\alpha$ and $\beta$ mean Cartesian coordinates and are assumed to obey the Einstein summation convention.

The parameter $v$ is a constant which describes the excluded volume of segments, and $V_{\alpha\beta\alpha^\prime\beta^\prime}$ is a 4th rank tensor which describes the strength of the nematic interaction.
We will analyze the 2nd order term $F_2$ in the GL expansion derived from the Hamiltonian eq.(\ref{eqn:Hamiltonian}) to judge whether the spinodal decomposition occurs or not.
The explicit expression of $F_2$ can be written as
\begin{align}
F_2&=\frac{k_{\rm B}T}{2}\int d\mbox{\boldmath$r$}_1d\mbox{\boldmath$r$}_2C^{-1}(\mbox{\boldmath$r$}_1- \mbox{\boldmath$r$}_2) \delta\phi(\mbox{\boldmath$r$}_1)\delta\phi(\mbox{\boldmath$r$}_2)
+\frac{k_{\rm B}T}{2}\int d\mbox{\boldmath$r$}_1d\mbox{\boldmath$r$}_2D_{\alpha\beta\alpha\beta}^{-1}(\mbox{\boldmath$r$}_1-\mbox{\boldmath$r$}_2) S_{\alpha\beta}(\mbox{\boldmath$r$}_1)S_{\alpha\beta}(\mbox{\boldmath$r$}_2),\label{eqn:linear_term}
\end{align}
where $C^{-1}(\mbox{\boldmath$r$}_1-\mbox{\boldmath$r$}_2)$ and $D_{\alpha\beta\alpha^\prime\beta^\prime}^{-1}(\mbox{\boldmath$r$}_1-\mbox{\boldmath$r$}_2)$ are inverse functions of the density correlation function and orientation correlation function in the uniform melt, respectively\cite{Leibler}.
These correlation function are defined as
\begin{align}
C(\mbox{\boldmath$r$}_1-\mbox{\boldmath$r$}_2)&=\left\langle \delta\hat{\phi}(\mbox{\boldmath$r$};\Gamma)\delta\hat{\phi}(\mbox{\boldmath$r$}^\prime;\Gamma)  \right\rangle,\\
D_{\alpha\beta\alpha^\prime\beta^\prime}(\mbox{\boldmath$r$}_1-\mbox{\boldmath$r$}_2)&=\left\langle \hat{S}_{\alpha\beta}(\mbox{\boldmath$r$};\Gamma)\hat{S}_{\alpha^\prime\beta^\prime}(\mbox{\boldmath$r$}^\prime;\Gamma)  \right\rangle,
\end{align}
where the notation $\langle\cdots \rangle$ means the physical quantities $\cdots$ averaged over the uniform melt including the interaction between segments.
The inverse function denoted by $f^{-1}(\mbox{\boldmath$r$})$ is defined as
\begin{align}
\int d\mbox{\boldmath$r$}^\prime f(\mbox{\boldmath$r$}-\mbox{\boldmath$r$}^\prime)f^{-1}(\mbox{\boldmath$r$}^\prime-\mbox{\boldmath$r$}^{\prime\prime})=\delta(\mbox{\boldmath$r$}-\mbox{\boldmath$r$}^{\prime\prime}).
\end{align}

In the analysis of eq.(\ref{eqn:linear_term}), there are two steps.
The first step is to obtain $C(\mbox{\boldmath$r$})$ and $D_{\alpha\beta\alpha^\prime\beta^\prime}(\mbox{\boldmath$r$})$ based on the information of an ideal melt.
To perform this first step, we use the random phase approximation(RPA)\cite{Leibler}.
The second step is to introduce the coupling between the density and the orientation. Such a coupling should be at least third order in the fluctuations because the density and the orientation fluctuations are decoupled up to 2nd order in the GL expansion due to the symmetry of the reference uniform state. 
As we have discussed in Sec.\ref{sec:introduction}, the result of X-ray scattering experiments and MD simulations suggest that the SD is originating from the coupling between density and orientation.
As will be shown later, an elimination of the degree of freedom of the orientation resolves this problem.

In the following sections \ref{sec:transfer_matrix}, \ref{sec:RPA}, and \ref{sec:coupling}, we discuss the detail of the method to perform the above first and second steps. 
We describe the statistical nature of a single ideal chain in terms of transfer matrix which includes the information of the stiffness of the polymer chain.
Then, by eliminating the degree of freedom of the orientation, we clarify how the nematic interaction affects the density fluctuations.
Finally, we calculate the spatial correlation function of the density fluctuations.

\begin{figure}[H]
\begin{center}
\includegraphics[clip, width=5.0cm]{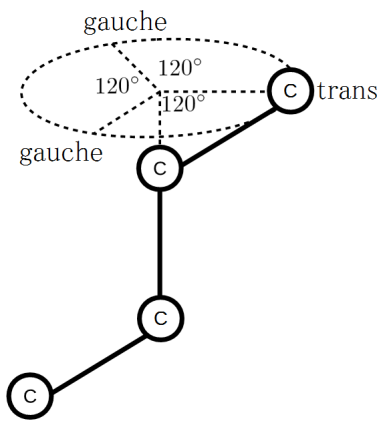}
\caption{3 possible local conformations of the dihedral angle between consecutive three bonds.
The most stable conformation is called ``trans conformation''(dihedral angle $0^\circ$), and the metastable ones are called ``gauche conformations''(dihedral angle $\pm 120^\circ$), respectively.}
\label{fig:transgauche}
\end{center}
\end{figure}

\begin{figure}[H]
\begin{center}
\includegraphics[clip, width=4.0cm]{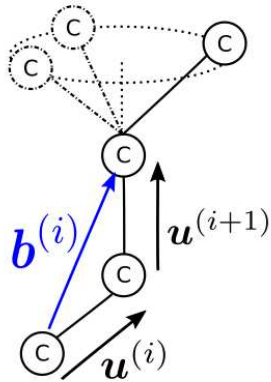}
\caption{The black vectors (denoted as $\mbox{\boldmath$u$}$) are the original bond vectors, while the blue one (denoted as $\mbox{\boldmath$b$}$) is the coarse-grained bond vector.
The index $i$ specifies the bond.}
\label{fig:cg}
\end{center}
\end{figure}

\subsection{Single chain statistics}\label{sec:transfer_matrix}
We consider a polymer chain where the backbone atoms are sequentially connected by $N+1$ covalent bonds.
The total number of possible states of the conformation is enormous, which originates from the rotation of the bonds.
However, these is a certain restriction on the rotation of the bonds.
As is shown in Fig.\ref{fig:transgauche}, the dihedral angle composed of consecutive 3 bonds takes either $0^\circ$ or $\pm 120^\circ$.
The conformation with angle $0^\circ$ is the most stable conformation, which is called ``trans conformation''.
On the other hand, the angles $\pm 120^\circ$ are metastable and called ``gauche conformations''.
Although the energies of two gauche conformations are in general different due to the difference in the side atomic groups, we neglect such a difference and assume that the two gauche conformations are degenerating for simplicity.
\begin{figure}[h]
\begin{center}
\includegraphics[clip, width=8.0cm]{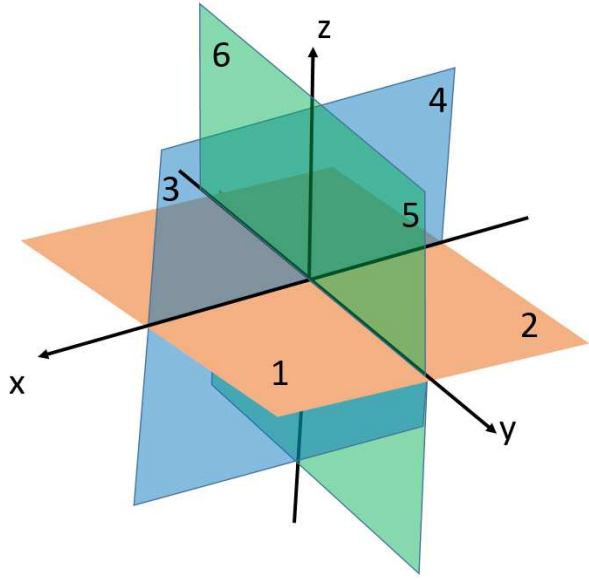}
\caption{A coarse-grained bond vector can take 12 different states (directions pointing to the 12 face center positions from the origin).
We identify these vectors by using an index $\mu=1, 2, \cdots, 12$.
The vectors $\mu = 1, 2, \cdots 6$ are shown in the figure.
The vectors $\mu = 7, \cdots,12$ are defined as the vectors pointing to the opposite directions from the vectors $\mu = 1, \cdots,6$, respectively.}
\label{fig:surface}

\end{center}
\end{figure}

When we specify the directions of the first and the second bonds of a single chain (see Fig.~\ref{fig:cg}), the subsequent bond vectors are always on a diamond lattice spanned by the first 2 bonds, {\it i.e.}
the bond vector $\mbox{\boldmath$u$}$ can be chosen as one of the 8 primitive translational vectors on the lattice: $\pm( -u, -u, -u)/\sqrt{3}$, $\pm( -u, u, u)/\sqrt{3}$, $\pm(u,  -u,  u)/\sqrt{3}$, $\pm(u, u,  -u)/\sqrt{3}$ where $u$ is the length of the bond.
When we call the covalent bond connecting backbone atoms an ``original bond'', we define the coarse-grained bond as a sum of consecutive two original bonds(shown by blue arrow in Fig.~\ref{fig:cg}).
By this definition, we understand that the coarse-grained bond vector $\mbox{\boldmath$b$}$ points one of the face center positions from the origin as is shown in Fig.~\ref{fig:surface}.
Thus, the coarse-grained bond vector can be described as $(\pm b, \pm b, 0), (\pm b, 0, \pm b), (0, \pm b, \pm b)$ in no particular order(12 vectors shown in Fig.~\ref{fig:surface}).
In the following, we denote the coarse-grained bond vector as $\mbox{\boldmath$e$}^{(\mu)}$ where the superscript $\mu$ specifies the direction of the coarse-grained bond (each index corresponds to the surface number in Fig.~\ref{fig:surface}), for example,  $\mbox{\boldmath$e$}^{(1)}=(b,b,0)$ and so on. 
As the original bond cannot freely rotate, a constraint condition is imposed on the coarse-grained bond vectors.
Depending on the trans and two gauche conformations, we understand that the angle between 2 consecutive coarse-grained bond vectors can take either $0^\circ$(corresponding to trans conformation in the original bonds) or $60^\circ$(corresponding to gauche).
In the original bonds, there are 2 different gauche conformations, while in coarse-grained bond vectors, the number of gauche conformations is 4. 
As we will only use the coarse-grained bond in the following, hereafter we will refer the coarse-grained bond as simply 'bond' unless otherwise noted.

In general, the trans conformation has lower energy than the gauche one, and therefore the following relation holds: 
\begin{equation}
\Delta\varepsilon \equiv \varepsilon_{\rm gauche}-\varepsilon_{\rm trans} \ge 0,
\end{equation}
where $\varepsilon_{\rm trans}$ and $\varepsilon_{\rm gauche}$ are the energies of the trans and the gauche conformations for the bonds.
This parameter $\Delta\varepsilon$ determines the statistical properties of the chain conformation.  When the two bond vectors are parallel(trans case), the statistical weight of this state is assumed to be $1$ while the statistical weight for the state with angle $60^{\circ}$(gauche case) is $\delta$ defined by
\begin{align}
\delta&=\exp{\left(-\beta \Delta\varepsilon \right)},
\end{align}
where $\beta=1/(k_{\rm B}T)$, $k_{\rm B}$ is Boltzman constant and $T$ is the temperature.
The statistical weights for the other states are $0$ because they are forbidden.
These constraints on the conformation can be described in terms of the following transfer matrix,
\begin{equation}
\mathcal{T}=\left(
\begin{array}{cccccccccccc}
1& 0& \delta& 0& \delta& 0& 0& 0& 0& \delta& 0& \delta\\
0& 1& 0& \delta& \delta& 0& 0& 0& \delta& 0& 0& \delta\\
\delta& 0& 1& 0& \delta& \delta& 0& \delta& 0& 0& 0& 0\\
0& \delta& 0& 1& \delta& \delta& \delta& 0& 0& 0& 0& 0\\
\delta& \delta& \delta& \delta& 1& 0& 0& 0& 0& 0& 0& 0\\
0& 0& \delta& \delta& 0& 1& \delta& \delta& 0& 0& 0& 0\\
0& 0& 0& \delta& 0& \delta& 1& 0& \delta& 0& \delta& 0\\
0& 0& \delta& 0& 0& \delta& 0& 1& 0& \delta& \delta& 0\\
0& \delta& 0& 0& 0& 0& \delta& 0& 1& 0& \delta& \delta\\
\delta& 0& 0& 0& 0& 0& 0& \delta& 0& 1& \delta& \delta\\
0& 0& 0& 0& 0& 0& \delta& \delta& \delta& \delta& 1& 0\\
\delta& \delta& 0& 0& 0& 0& 0& 0& \delta& \delta& 0& 1
\end{array}
\right).
\label{eqn:transfermatrix}
\end{equation}
The ${\mu\nu}-$component of the transfer matrix $\mathcal{T}$ means the statistical weight of the local conformation where the direction of $i$-th bond vector is parallel to $\mbox{\boldmath$e$}^{(\mu)}$ and $(i+1)$-th bond vector is parallel to $\mbox{\boldmath$e$}^{(\nu)}$, respectively.
For example, $\mathcal{T}_{11}$ is the statistical weight of the trans conformation where both $i-$th and $(i+1)$-th bond vectors are parallel to $\mbox{\boldmath$e$}^{(1)}$.
In this case $\mathcal{T}_{11}=1$.

By using this transfer matrix we can calculate the average of physical quantities of a single chain.
The partition function of a single chain is obtained as
\begin{align}
Z(T,N)&=\sum_{\mu=1}^{12}\sum_{\nu=1}^{12}\left(\mathcal{T}^N  \right)_{\mu\nu}\nonumber\\
&=12(1+4\delta)^{N-1},
\end{align}
where $(\mathcal{T}^N)_{\mu\nu}$ describes the $\mu\nu$-component of $\mathcal{T}^N$.
The prefactor 12 means that the initial bond is chosen from 12 possible vectors.
The factor $1+4\delta$ comes from the fact that the 2 consecutive bonds can take 1 trans and 4 gauche conformations.
To calculate the physical quantities for a single chain, we introduce a path integral $Q(0, \mu, \mbox{\boldmath$r$};N,\nu,\mbox{\boldmath$r$}^\prime)$ which describes the statistical weight of the event where the $0$-th bond is found at the position $\mbox{\boldmath$r$}$ and is parallel to the vector $\mbox{\boldmath$e$}^{(\mu)}$ and the $N$-th bond is at the position $\mbox{\boldmath$r$}^\prime$ and is parallel to the vector $\mbox{\boldmath$e$}^{(\nu)}$, respectively.
Then, the path integral can be written as:
\begin{align}
Q(0, \mu, \mbox{\boldmath$r$};N, \nu, \mbox{\boldmath$r$}^\prime)&=\sum_{\left\{\mbox{\boldmath$b$}^{(i)}\right\}}
(\mathcal{T}^N)_{\mu \nu} \delta\left(\mbox{\boldmath$r$}^\prime+\frac{\mbox{\boldmath$b$}^{(N)}}{2}-\left(\mbox{\boldmath$r$}+\frac{\mbox{\boldmath$b$}^{(0)}}{2}+\sum_{i=1}^N \mbox{\boldmath$b$}^{(i)}\right)\right)\nonumber\\
&=\int d\mbox{\boldmath$q$}\left(\tilde{\mathcal{T}}^N(\mbox{\boldmath$q$})\right)_{\mu\nu} \exp{\left[i \mbox{\boldmath$q$}\cdot\left(\mbox{\boldmath$r$}^\prime-\mbox{\boldmath$r$}\right)\right]}, 
\end{align}
where 
\begin{align}
\left(\tilde{\mathcal{T}}(\mbox{\boldmath$q$})\right)_{\mu \nu}&=\exp{\left(-i \mbox{\boldmath$q$}\cdot \frac{\mbox{\boldmath$e$}^{(\mu)}}{2} \right)}\mathcal{T}_{\mu \nu}\exp{\left(-i \mbox{\boldmath$q$}\cdot \frac{\mbox{\boldmath$e$}^{(\nu)}}{2} \right)}.
\end{align}
\hspace{3mm}We can compute the physical quantities of the single chain by using the above path-integral combined with the transfer matrix defined in eq.~(\ref{eqn:transfermatrix}).
For example the segment density field $\phi(\mbox{\boldmath$r$})$ and the orientation field $S_{\alpha\beta}(\mbox{\boldmath$r$})$ are given by
\begin{align}
\delta\phi(\mbox{\boldmath$r$})&=\langle \delta\hat{\phi}(\mbox{\boldmath$r$}; \Gamma) \rangle_0 = \left\langle \sum_{i=0}^N \delta(\mbox{\boldmath$r$}-\mbox{\boldmath$r$}^{(i)})\right\rangle_0 - \bar{\phi}, \label{eqn:density}\\
S_{\alpha\beta}(\mbox{\boldmath$r$})&=\langle \hat{S}_{\alpha\beta}(\mbox{\boldmath$r$}; \Gamma) \rangle_0 =\left\langle \sum_{i=0}^N\frac{3}{2}\left(b_{\alpha}^{(i)}b_{\beta}^{(i)} -\frac{1}{3}\delta_{\alpha\beta} \right) \delta(\mbox{\boldmath$r$}-\mbox{\boldmath$r$}^{(i)}) \right\rangle_0, \label{eqn:orientation}
\end{align}
where the notation $\langle \cdots \rangle_0$ means the average quantities over the ensemble of the set of the ideal chains.

Then we can obtain the correlation function of the segment density and the orientation of the single chain, respectively, as
\begin{align}
C^{(0)}(\mbox{\boldmath$r$}- \mbox{\boldmath$r$}^\prime)&=\langle \delta\hat{\phi}(\mbox{\boldmath$r$};\Gamma)\delta\hat{\phi}(\mbox{\boldmath$r$}^\prime ;\Gamma)  \rangle_0 , \label{eqn:density_corre} \\
D_{\alpha\beta\alpha^\prime\beta^\prime}^{(0)}(\mbox{\boldmath$r$}- \mbox{\boldmath$r$}^\prime)&=\langle \hat{S}_{\alpha\beta}(\mbox{\boldmath$r$};\Gamma)\hat{S}_{\alpha^\prime\beta^\prime}(\mbox{\boldmath$r$}^\prime;\Gamma)  \rangle_0, \label{eqn:orientation_corre}
\end{align}
where superscript $(0)$ on the left hand sides of eqs.(\ref{eqn:density_corre}) and (\ref{eqn:orientation_corre}) means that the quantities are evaluated for the set of the ideal chains.
It should be noted that the cross correlation function, $\langle\delta\hat{\phi}(\mbox{\boldmath$r$};\Gamma)\hat{S}_{\alpha\beta}(\mbox{\boldmath$r$}^\prime;\Gamma) \rangle_0$, up to 2nd order vanishes due to the symmetry.

\subsection{Random phase approximation(RPA)}\label{sec:RPA}
We express using RPA the 2nd order terms in the GL expansion for the polymer melt system, where the coefficients in eq.~(\ref{eqn:linear_term}) are described by the correlation functions eq.~(\ref{eqn:density_corre}) and (\ref{eqn:orientation_corre}) for the ideal chain system.
The RPA procedure can easily be performed in the Fourier space.
The Fourier transformation of an arbitrary function of the position, $f(\mbox{\boldmath$r$})$, is defined as,
\begin{align}
\tilde{f}(\mbox{\boldmath$q$})&=\int d\mbox{\boldmath$r$} f(\mbox{\boldmath$r$}) \exp{\left[-i \mbox{\boldmath$q$}\cdot \mbox{\boldmath$r$} \right]}.
\end{align}
The second order term in the expansion of $F$ is written in the Fourier space as
\begin{align}
F_2[\delta \tilde{\phi}(\mbox{\boldmath$q$}), \tilde{S}_{\alpha\beta}(\mbox{\boldmath$q$})]&=\frac{1}{2}\int d\mbox{\boldmath$q$}\frac{N^{-1}\tilde{C}^{(0)}(\mbox{\boldmath$q$})}{1+\beta v N^{-1}\tilde{C}^{(0)}(\mbox{\boldmath$q$})}|\delta\tilde{\phi}(\mbox{\boldmath$q$})|^2 \nonumber\\
&+\frac{1}{2}\int d\mbox{\boldmath$q$}\frac{N^{-1}\tilde{D}^{(0)}(\mbox{\boldmath$q$})}{1+\beta V N^{-1}\tilde{D}^{(0)}(\mbox{\boldmath$q$})}\tilde{S}_{\alpha\beta(\mbox{\boldmath$q$})}\tilde{S}_{\beta\alpha(-\mbox{\boldmath$q$})}\label{eqn:free_energy},
\end{align}
where $V$ and $\tilde{D}^{(0)}$ are described by the components of the nematic parameter $V_{\alpha\beta\alpha^\prime\beta^\prime}$ and the power spectrum of the orientation field $\tilde{D}_{\alpha\beta\alpha^\prime\beta^\prime}^{(0)}$ defined in \ref{sec:symmetry}, respectively, and the factor $N^{-1}$ is a normalization factor.
In the derivation of eq.(\ref{eqn:free_energy}), we used the fact that $\tilde{D}_{\alpha\beta\alpha^\prime\beta^\prime}^{(0)}$ and $V_{\alpha\beta\alpha^\prime\beta^\prime}$ are isotropic tensors due to the symmetry of the reference uniform state.
As the segment density in the one component uniform melt is characterized by the repulsive interaction $v>0$ which is defined in eq.(\ref{eqn:Hamiltonian}), an effective attraction is necessary for reproducing the SD.
We notice that the density and the orientation are decoupled in the terms of the free energy up to 2nd order, which means that the segment density by itself cannot induce an instability in the reference uniform state. 
However, the experimental result implies that the density fluctuation leads to the instability of the uniform state, which is a sign of the SD.
In order to reproduce such experimental results, we should introduce the effective attraction between segments into our model.

\subsection{The relationship between density and orientation}\label{sec:coupling}
As we discussed in Sec.\ref{sec:introduction}, the X-ray scattering experiments and MD simulations on the early stage of the crystallization suggested that the coupling between density and orientation is important for SD.

As a non-conserved order parameter (for example orientation) relaxes faster than a conserved one (e.g. density) in general, we can eliminate the degree of freedom of the orientation compared with the density.
Then we integrate the partition function over only the degrees of freedom of the orientation as follows:
\begin{align}
Z&=\int d\Gamma \exp(-\beta \mathcal{\hat{H}}-\beta \hat{\mathcal{W}}_S[\hat{S}_{\alpha\beta}(\mbox{\boldmath$r$})]-\beta \hat{\mathcal{W}}_\phi[\delta\hat{\phi}(\mbox{\boldmath$r$})])\nonumber\\
&=\int d\mbox{\boldmath$r$}^{(0)}d\mbox{\boldmath$b$}^{(0)}d\mbox{\boldmath$b$}^{(1)}\cdots d\mbox{\boldmath$b$}^{(N)} \exp(-\beta \mathcal{\hat{H}}_0)\left(1-\beta \hat{\mathcal{W}}_S[\hat{S}_{\alpha\beta}(\mbox{\boldmath$r$})]-\beta \hat{\mathcal{W}}_\phi[\delta\hat{\phi}(\mbox{\boldmath$r$})] \right)\nonumber\\
&=\int d\mbox{\boldmath$r$}^{(0)} d\mbox{\boldmath$r$}^{(1)} \cdots d\mbox{\boldmath$r$}^{(N)}d\mbox{\boldmath$b$}^{(0)}d\mbox{\boldmath$b$}^{(1)}\cdots d\mbox{\boldmath$b$}^{(N)} \delta\left(\mbox{\boldmath$r$}^{(N)}+\frac{\mbox{\boldmath$b$}_N}{2}-\left(\mbox{\boldmath$r$}^{(0)}+\frac{\mbox{\boldmath$b$}^{(0)}}{2}+\sum_{i=1}^N \mbox{\boldmath$b$}^{(i)}\right) \right)\nonumber\\
&\times \exp(-\beta \mathcal{\hat{H}}_0)\left(1-\beta \hat{\mathcal{W}}_S[\hat{S}_{\alpha\beta}(\mbox{\boldmath$r$})]-\beta \hat{\mathcal{W}}_\phi[\delta\hat{\phi}(\mbox{\boldmath$r$})] \right)\nonumber\\
&=\int d\mbox{\boldmath$r$}^{(0)} d\mbox{\boldmath$r$}^{(1)} \cdots d\mbox{\boldmath$r$}^{(N)} \exp{\left(-\beta \mathcal{\hat{H}}_0^\prime \right)}\left(1-\beta \hat{\mathcal{W}}_\phi^\prime[\delta\hat{\phi}(\mbox{\boldmath$r$})] \right).
\end{align}
Here, we define $\hat{W}_\phi^\prime[\delta\hat{\phi}(\mbox{\boldmath$r$})]$ and $v^\prime$ as
\begin{align}
\hat{\mathcal{W}}_\phi^\prime[\delta\hat{\phi}(\mbox{\boldmath$r$})]&=\frac{v^\prime}{2}\int d\mbox{\boldmath$r$} \left\{\delta\hat{\phi}(\mbox{\boldmath$r$})  \right\}^2\\
v^\prime&=v_{\rm ex}+v_{\rm ne}(l_{\rm p})\label{eqn:contact_interaction},
\end{align}
where $v_{\rm ex}$ and $v_{\rm ne}$ are the excluded volume and the effect of the nematic interaction, and $l_{\rm p}$ is the persistence length defined as
\begin{align}
l_{\rm p}&=b\exp{\left[\beta \Delta\varepsilon \right]}.
\end{align}
The parameter $v_{\rm ne}$ can be obtained by using a perturbation theory and pre-averaging approximation as 
\begin{align}
v_{\rm ne}&=\frac{9}{4}\frac{V}{N}\left[\frac{1}{\displaystyle{1-\exp{\left[\frac{3}{l_{\rm p}}  \right]}}}\left(\frac{\displaystyle{1-\exp{\left[-3\frac{N-1}{l_{\rm p}}  \right]}}}{\displaystyle{1-\exp{\left[-\frac{3}{l_{\rm p}}  \right]}}}-N \right) \right]-\frac{3V}{4}.
\end{align}
The detail of this calculation is shown in \ref{sec:detail_of_calculation}.

\begin{figure}[H]
\begin{center}
\includegraphics[width=10cm]{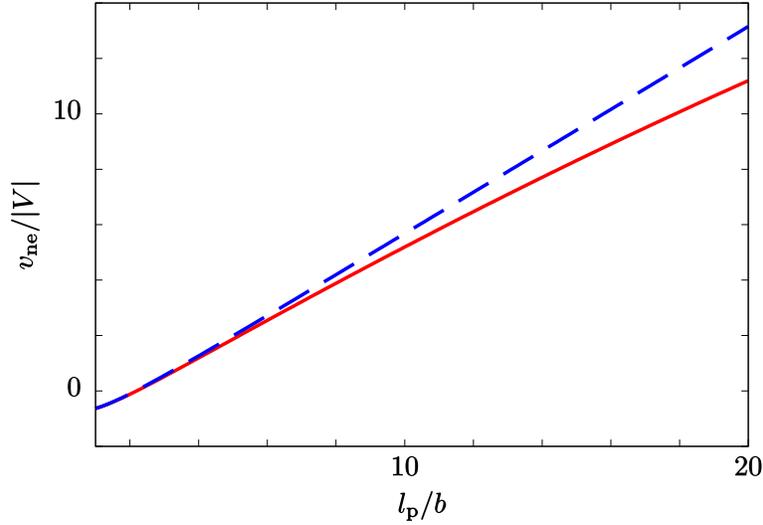}
\caption{The vertical axis is $v_{\rm ne}/| V |$ and the horizontal axis is $l_{\rm p}/b$ for $N=51$(red solid line) and $N\rightarrow \infty$(blue dashed line).}
\label{fig:mean}
\end{center}
\end{figure}

In our model, $V(<0)$, which implies that the system prefers nematic order, leads to $v_{\rm ne}<0$ for $l_{\rm p}/b\geq 3/\ln{4}$ for $N\rightarrow \infty$ as shown in Fig.~\ref{fig:mean}.
According to the behavior of $v_{\rm ne}$, the nematic interaction decreases the excluded volume for the case with $l_{\rm p}/b\geq 3/\ln{4}$ and large $|V|$, where the parameter $v_{\rm ne}(\propto V)$ can induce the effective attraction by overcoming the excluded volume parameter $v$.

Let us summarize the results shown in this subsection.
By eliminating the degree of freedom of the orientation, the effect of the nematic order on the density field is formulated.
Depending on the persistence length $l_{\rm p}$ and the nematic parameter $V$, the effect of the nematic order can play the role of an effective attraction between the segments. 
As the attraction leads to an instability of the reference state in RPA procedure, the SD can be induced by the coupling between the orientation and the density.

\begin{figure}[H]
\begin{center}
\includegraphics[clip, width=10.0cm]{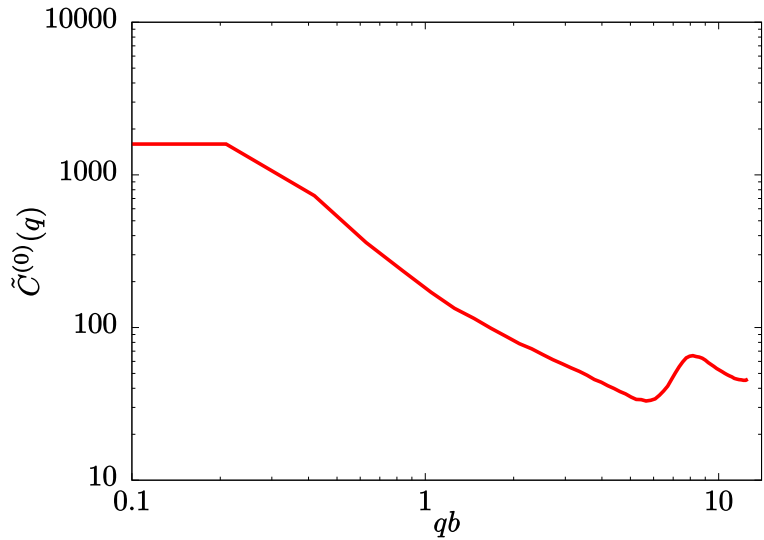}
\caption{The power spectrum of the segment density vs. wave number, where the typical energy for $\Delta\varepsilon$ is $2.5 \times 10^3[{\rm J}/{\rm mol}]$\cite{Strobl}, $N=51$ and $T=400$[K].}
\label{fig:correlation0}
\end{center}
\end{figure}

\begin{figure}[H]
\begin{center}
\includegraphics[clip, width=10.0cm]{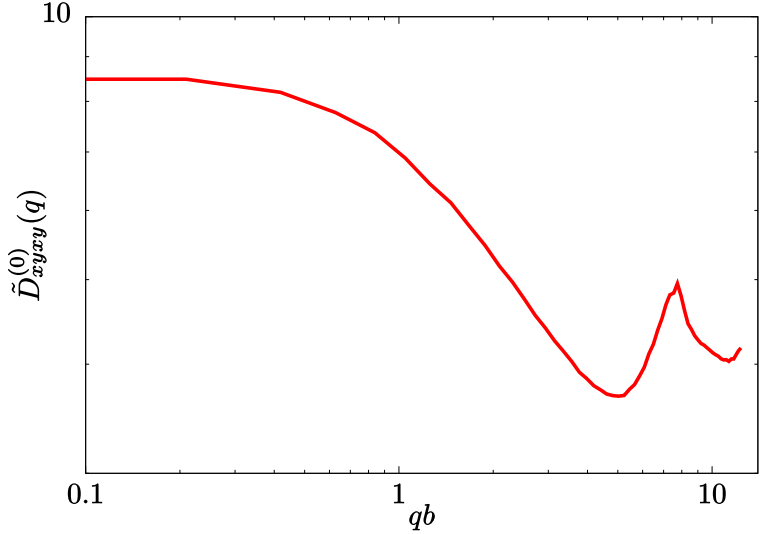}
\caption{The power spectrum of the segment orientation vs. wave number.
The parameters used in the calculation are the same as those in Fig.~\ref{fig:correlation0}.
}
\label{fig:orientation_correlation0}
\end{center}
\end{figure}

\section{Result and discussion}\label{sec:result}
\subsection{Correlation functions of an ideal chain}
\hspace{3mm}The power spectrum of the segment density and that of the orientation of the segments are shown in Figs.~\ref{fig:correlation0} and \ref{fig:orientation_correlation0}, respectively.
In these figures, we calculate these power spectra for the case $N=51$ and $\beta \Delta\varepsilon\simeq 0.76$ (corresponding to $2.5\times10^3[{\rm J}/{\rm mol}]$ at $400$[K]) which is a representative value obtained by Raman scattering experiments\cite{Strobl}.

Figure~\ref{fig:correlation0} shows the relationship between the magnitude of the wave number vector $q=\sqrt{q_x^2+q_y^2+q_z^2}$ and the Fourier component of the correlation function of the segment density $\tilde{C}^{(0)}(q)$.
When we calculated $\tilde{C}^{(0)}(q)$, we averaged the orientation of the initial bond vector over the isotropic distribution.
The behavior of $\tilde{C}^{(0)}$ is explained as follows.
$\tilde{C}^{(0)}$ is well fitted by the Debye function in the region $qb\ll2\pi$\cite{Doi_Edwards}.
On the other hand, we can recognize a structure of the single chain in the large wave number region (short length scale) in Fig.~\ref{fig:correlation0}, {\it i.e.} a peak around $qb\sim 8$.
This peak is due to the correlation between consecutive bonds in the microscopic scale.

The power spectrum of the orientation order parameter is shown in Fig.~\ref{fig:orientation_correlation0}, where the behavior is almost the same as that for the segment density.
This power spectrum of the orientation does not affect the following discussion because we consider the behavior of only the power spectrum of the density.
\subsection{Effective attraction derived from nematic interaction and SD}
In our model, density and orientation are coupled through the excluded volume and the nematic interactions.
The nematic interaction leads to a decrease in the excluded volume for the stiff polymer system, which means an effective attraction.
We prove that the strength of this attraction depends on the stiffness of the polymer chain (see Fig.~\ref{fig:mean}).
According to Fig.~\ref{fig:mean}, the strength of the attraction increases with the increase in the stiffness of the polymer chain.
In the situation where the effect of the chain ends can be neglected ({\it i.e.} $N\rightarrow \infty$), $v_{\rm ne}/|V|$ increases linearly with $l_{\rm p}$.
The ratio $v_{\rm ne}/|V|$ for the finite chain length shows us the different behavior from those of the infinite chain length due to the finite size effect.  

\begin{figure}[H]
\begin{center}
\includegraphics[clip, width=10cm]{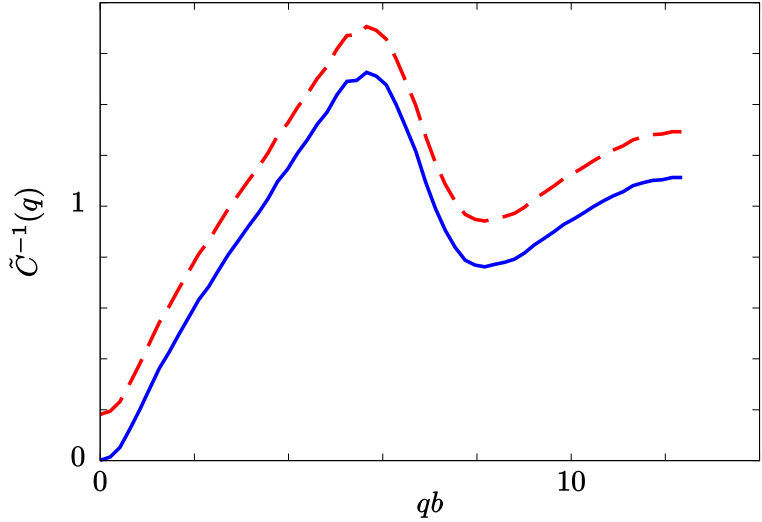}
\caption{The vertical axis is $\tilde{C}^{-1}(\mbox{\boldmath$q$})$ and the horizontal axis is $q b$.
Red(dashed) line describes the inverse of the power spectrum of density for $\beta v_{\rm ex}\simeq {0.20}$ and ${\beta V\simeq -0.10}$ and blue(solid) line describes the one for $\beta v_{\rm ex}\simeq {0.060}$ and ${\beta V\simeq -0.10}$.
}
\label{fig:spinodal}
\end{center}
\end{figure}

By substituting $v=v_{\rm ex}+v_{\rm ne}$ which are defined in eq.~(\ref{eqn:contact_interaction}), we can rewrite the power spectrum of the density fluctuation in eq.~(\ref{eqn:free_energy}) as
\begin{align}
\tilde{C}(\mbox{\boldmath$q$})&=\frac{N^{-1}\tilde{C}^{(0)}(\mbox{\boldmath$q$})}{1+N^{-1}(\beta v_{\rm ex} +\beta v_{\rm ne})\tilde{C}^{(0)}(\mbox{\boldmath$q$})}\label{eqn:condition_for_instability}.
\end{align}
The condition of the instability shown in Fig.~\ref{fig:spinodal} is $\beta v_{\rm ex}{\simeq 0.02}$, $\beta {V\simeq -0.2}$, and $N=51$.
The effective attraction derived from the nematic interaction induces the instability of the uniform melt.
To induce this instability, we should note that the excluded volume parameter $\beta v_{\rm ex}$ and the effective attraction $\beta v_{\rm ne}$ satisfy the condition as
\begin{align}
\beta v_{\rm ex}+\beta v_{\rm ne}&=-\frac{N}{\tilde{C}^{(0)}(|\mbox{\boldmath$q$}|=0)}.
\end{align}
The increase in the parameters $v_{\rm ex}$ and $V$ corresponds to a deeper quench.
The detail of this increase is expected to depend on the shape of the monomer or the interaction in the microscopic scale, which corresponds to the difference in the samples in experimental system.
This result is consistent with the experimental fact that the SD occurs in relatively deep quench case.

This instability leads to an increase in the correlation length of the density fluctuation in the early stage of the SD.
To estimate such correlation length, we use Cahn's linearized theory.
The starting point of this theory is the equation of continuity for the density field $\phi$ as
\begin{align}
\frac{\partial \delta\phi(\mbox{\boldmath$r$}, t)}{\partial t}&=L \nabla^2 \frac{\delta F}{\delta (\delta\phi(\mbox{\boldmath$r$}))}\label{eqn:continuity},
\end{align}
where $L$ is the kinetic coefficient.
In Cahn's linearized theory, the free energy $F$ in eq.(\ref{eqn:continuity}) is approximated as the 2nd order term of GL expansion, $F_2$.
Fourier transform of eq.~(\ref{eqn:continuity}) and expanding the power spectrum of the density fluctuation around $q=0$ lead to the expression as 
\begin{align}
\frac{\partial\delta\tilde{\phi}(\mbox{\boldmath$q$}, t)}{\partial t}&=-L q^2\left(\left. \tilde{C}^{-1}(\mbox{\boldmath$q$})\right|_{\mbox{\boldmath$q$}\rightarrow \mbox{\boldmath$0$}+}+\frac{1}{2}\left. \frac{\partial^2 \tilde{C}^{-1}(\mbox{\boldmath$q$})}{\partial \mbox{\boldmath$q$}\mbox{\boldmath$q$}}\right|_{\mbox{\boldmath$q$}\rightarrow \mbox{\boldmath$0$}+} :\mbox{\boldmath$q$}\mbox{\boldmath$q$} \right) \delta\tilde{\phi}(\mbox{\boldmath$q$}). 
\end{align}
We can solve this equation to obtain 
\begin{align}
\delta \phi(q) &\sim \exp{\left[\lambda(q)t \right]}\\
\lambda(q)&=-L q^2\left(\left. \tilde{C}^{-1}(\mbox{\boldmath$q$})\right|_{\mbox{\boldmath$q$}\rightarrow \mbox{\boldmath$0$}+}+\frac{1}{2}\left. \frac{\partial^2 \tilde{C}^{-1}(\mbox{\boldmath$q$})}{\partial \mbox{\boldmath$q$}^2}\right|_{\mbox{\boldmath$q$}\rightarrow \mbox{\boldmath$0$}+} \mbox{\boldmath$q$}^2 \right)\nonumber\\
&=-L\left(c q^2 +\frac{1}{2}\kappa q^4 \right),
\end{align}
where
\begin{align}
c&=C^{-1}(q=0)=\frac{1}{N}+\beta v \label{eqn:c}\\
\kappa&=\left. -N\frac{\frac{\displaystyle{d^2}}{\displaystyle{d q^2}}C^{(0)}(q)}{\left\{C^{(0)}(q)\right\}^2}\right|_{q=0}\label{eqn:kappa}.
\end{align}
The wave number $q^{\ast}$ which gives the maximum of $\lambda(q)$ corresponds to the inverse of the correlation length calculated as:
\begin{align}
q^\ast&=\sqrt{\frac{-c}{\kappa}}\label{eqn:q_ast}
\end{align} 

\begin{figure}[H]
\begin{center}
\includegraphics[clip, width=10cm]{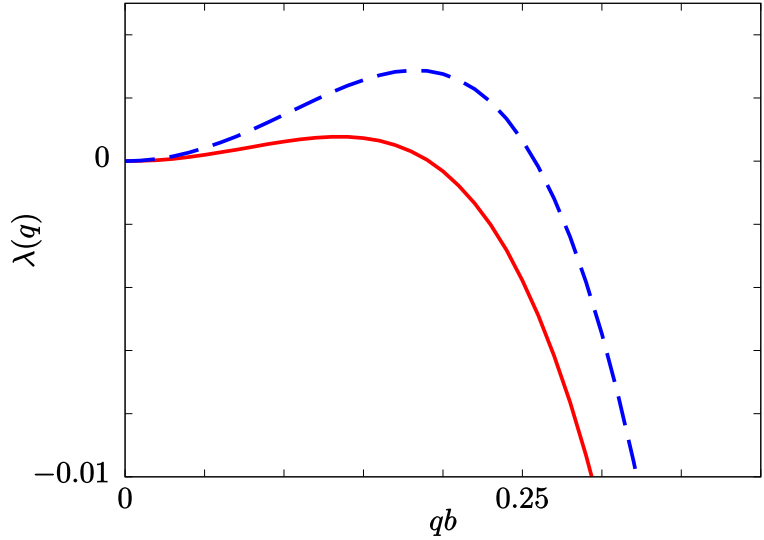}
\caption{The vertical axis is $\lambda(q)$ and the horizontal axis is $q b$ for $N=51$ where the contact interaction parameters are $\beta v\simeq 0.05$ and $\beta V\simeq -0.10$.
Red(solid) line describes $\lambda(q)$ for $\beta \Delta\varepsilon \simeq 1.51$(corresponding to $1.5\times3.3\times 10^{3}$[J/mol] at $400$[K] where $2.5\times 10^{3}-3.3\times 10^{3}$[J/mol] is determined by Ramann scattering on polyetylene\cite{Strobl}), and blue(dashed) line describes $\lambda(q)$ for $\beta\Delta\varepsilon \simeq 1.76$(corresponding to $1.75\times3.3\times10^3$[J/mol] at $400$[K]), respectively.}
\label{fig:lambda}
\end{center}
\end{figure}

We show the result of the growth rate of the density fluctuation in Fig. \ref{fig:lambda}.
This implies that the correlation length of the density fluctuation in stiffer polymer system is larger than that in the flexible polymer system.
As shown in the following, the competition between the excluded volume and the effective attraction determines the correlation length of the density fluctuation.
\begin{figure}[H]
\begin{center}
\includegraphics[clip, width=10cm]{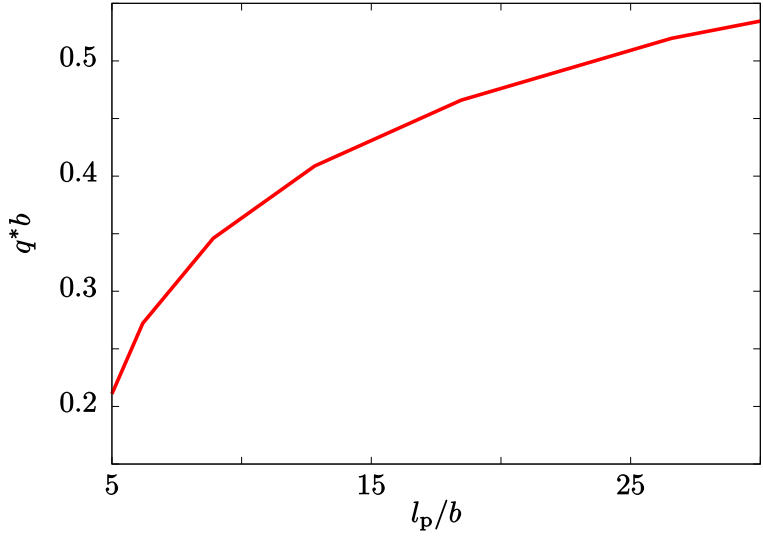}
\caption{The vertical axis is $q^{\ast} b$ and the horizontal axis is $l_{\rm p}/b$ for $N=51$, where the parameters $\beta v$ and $\beta V$ are the same as those used in Fig.~\ref{fig:lambda}.}
\label{fig:q_max}
\end{center}
\end{figure}
The relationship between $q^{\ast}$ and $l_{\rm p}$ shown in Fig.~\ref{fig:q_max} implies that the correlation length of the density fluctuation for the stiff polymer melt is smaller than that for the flexible polymer melt.

By Guinier approximation which describes the behavior of the structure factor around $|\mbox{\boldmath$q$}|=0$, we can obtain the 2nd order derivative of the power spectrum of the density at $|\mbox{\boldmath$q$}|=0$ as
\begin{align}
\left. \frac{d^2}{d q^2}C^{(0)}(q)\right|_{q=0}&=-\frac{N^2 R_{\rm g}^2}{3}\label{eqn:guiner},
\end{align}
where $R_{\rm g}$ is the gyration radius of the polymer chain\cite{Doi_Edwards}.
To substitute eqs.~(\ref{eqn:c}) and (\ref{eqn:kappa}) into eq.~(\ref{eqn:q_ast}) and using eq.~(\ref{eqn:guiner}), the expression of $q^\ast$ is written as 
\begin{align}
q^\ast&\sim \left(-\frac{\displaystyle{\frac{1}{N}+\beta v}}{\displaystyle{\frac{N^2 R_{\rm g}^2}{N^3}}}  \right)^{\frac{1}{2}}\nonumber\\
&=\frac{1}{R_{\rm g}}\left[N\left(|\beta v_{\rm ne}(l_{\rm p})|-\beta v_{\rm ex}\right)-1 \right]^\frac{1}{2}\label{eqn:expression_of_q_ast}.
\end{align}
The fact that the strength of the effective attraction depends on the persistence length of the polymer chain means that the nematic parameter, the excluded volume parameter and the stiffness of the polymer chain determine the characteristic wave number $q^\ast$.  
\section{Conclusion}\label{sec:conclusion}
By calculating GL expansion whose variables are the segment density field and the orientation of the segments around the uniform and isotropic reference state, we derived the condition for the occurrence of the SD in the induction period of the polymer crystallization.
The elimination of the degrees of freedom of the orientation leads to an effective attraction between segments depending on the given nematic parameter and the stiffness of the polymer chain expressed by the transfer matrix. 
As the stiffness of a polymer chain (corresponds to the parameter $\beta\Delta\varepsilon$ in our model) and the nematic order (corresponds to $V$) play the role of the effective attraction between segments, the SD in the induction period is observed in stiff polymer systems or in systems having larger nematic order. 
Introducing the characteristics of the chain conformation into our model ($\beta \Delta\varepsilon$ and $V$) determines the mechanism of the phase separation in the induction period.
This result agrees with the conclusion given by Chuang {\it et al.} who indicated that some parameters (for example, conformation of the initial state which depends on the stiffness of the polymer chain) determine the behavior in the induction period\cite{Chuang}. 

We can interpret these experimental results \cite{imai1} \cite{Z-G-Wang} in the context of the monomer shape.
Imai {\it et al.}\cite{imai1} reported the spinodal-assisted scenario by using experiments on the polyethylene terephthalate(PET), while Wang {\it et al.}\cite{Z-G-Wang} concluded that the SD does not occur in the induction period based on their experiments on isotactic polypropylene(iPP).
Intuitively, the monomer shape of PET is rod-like shape, which is expected to induce the strong nematic order.
On the other hand, iPP does not include the rod-like monomer, where the excluded volume is expected to be dominant in the induction period of the crystallization process.

Even though some authors interpreted the SD in the induction period as the result of an error in the X-ray scattering experiments\cite{Z-G-Wang}\cite{Panine}, our model indicates that the SD can occur depending on the stiffness of the polymer, the quench depth, and the shape of the monomer. 

In the future direction, the idea of the transfer matrix for a polymer chain is expected to be useful in explaining the dynamics of the polymer crystallization after the induction period, for example, the dynamics of the folding of the polymer chain and of the formation of the lamellar crystal.

\appendix
\def\thesection{Appendix \Alph{section}}

\section{Symmetry of the uniform melt}\label{sec:symmetry}
In this appendix, we show how to derive eq.~(\ref{eqn:free_energy}) on the basis of the isotropy of the reference uniform state, where the 4th rank tensor $V_{\alpha\beta\alpha^\prime\beta^\prime}$ should be an isotropic tensor.
Accordingly, the 4th rank tensors $\tilde{D}_{\alpha\beta\alpha^\prime\beta^\prime}^{(0)}$ and $\tilde{D}_{\alpha\beta\alpha^\prime\beta^\prime}$ are also isotropic tensors.
Then, $V_{\alpha\beta\alpha^\prime\beta^\prime}$, $\tilde{D}_{\alpha\beta\alpha^\prime\beta^\prime}^{(0)}$, and $\tilde{D}_{\alpha\beta\alpha^\prime\beta^\prime}$ can be generally expressed as
\begin{align}
V_{\alpha\beta\alpha^\prime\beta^\prime}&=V_1\delta_{\alpha\beta}\delta_{\alpha^\prime\beta^\prime}
+V_2\delta_{\alpha\alpha^\prime}\delta_{\beta\beta^\prime}+V_3\delta_{\alpha\beta^\prime}\delta_{\beta\alpha^\prime}
\label{eqn:V_iso}\\
\tilde{D}_{\alpha\beta\alpha^\prime\beta^\prime}^{(0)}&=\tilde{D}_1^{(0)}\delta_{\alpha\beta}\delta_{\alpha^\prime\beta^\prime}
+\tilde{D}_2^{(0)}\delta_{\alpha\alpha^\prime}\delta_{\beta\beta^\prime}+\tilde{D}_3^{(0)}\delta_{\alpha\beta^\prime}\delta_{\beta\alpha^\prime}
\label{eqn:D0_iso}\\
\tilde{D}_{\alpha\beta\alpha^\prime\beta^\prime}&=\tilde{D}_1\delta_{\alpha\beta}\delta_{\alpha^\prime\beta^\prime}
+\tilde{D}_2\delta_{\alpha\alpha^\prime}\delta_{\beta\beta^\prime}+\tilde{D}_3\delta_{\alpha\beta^\prime}\delta_{\beta\alpha^\prime}
\label{eqn:D_iso}.
\end{align}
It should be noted that $V_1$, $\tilde{D}_1^{(0)}$, and $\tilde{D}_1$ do not affect the free energy due to the traceless nature of the order parameter, $S_{\alpha\alpha}=0$.
Then, we rewrite the terms up to 2nd order in the order parameters in the GL expansion as 
\begin{align}
F_2&=\int d\mbox{\boldmath$q$} \tilde{C}^{-1}(\mbox{\boldmath$q$})|\delta \tilde{\phi}(\mbox{\boldmath$q$})|^2+\tilde{D}^{-1}(\mbox{\boldmath$q$})\tilde{S}_{\alpha\beta}(\mbox{\boldmath$q$})\tilde{S}_{\beta\alpha}(-\mbox{\boldmath$q$}),
\end{align}
where $\tilde{D}^{-1}=1/(\tilde{D}_2+\tilde{D}_3)$.
By using RPA and eqs.~(\ref{eqn:V_iso})-(\ref{eqn:D_iso}), we obtain the expression of $\tilde{D}^{-1}$ as 
\begin{align}
\tilde{D}^{-1}(\mbox{\boldmath$q$})&=\frac{\displaystyle{\tilde{D}_2^{(0)}+\tilde{D}_3^{(0)}}}{\displaystyle{1+\beta (V_2+V_3)(\tilde{D}_2^{(0)}+\tilde{D}_3^{(0)})}}\label{eqn:D_inverse_component}.
\end{align}
If we define $V$ and $\tilde{D}^{(0)}(\mbox{\boldmath$q$})$ as
\begin{align}
V&=V_2+V_3\\
\tilde{D}^{(0)}(\mbox{\boldmath$q$})&=\tilde{D}_2^{(0)}(\mbox{\boldmath$q$})+\tilde{D}_3^{(0)}(\mbox{\boldmath$q$}),
\end{align}
we can derive eq.~(\ref{eqn:free_energy}) by substituting the definition eqs.~(\ref{eqn:V_iso})-(\ref{eqn:D_iso}) into eq.~(\ref{eqn:D_inverse_component}).

\section{Relationship between density and nematic interaction}\label{sec:detail_of_calculation}
\subsection{Eliminating the degrees of freedom of orientation in nematic interaction}
In this appendix we show the detail of the calculation for eliminating the degrees of freedom of the orientation.
Starting point is the Hamiltonian eq.(\ref{eqn:Hamiltonian}) which includes the reference energy, the excluded volume interaction, and the nematic interaction:
\begin{align*}
\hat{\mathcal{H}}&=\hat{\mathcal{H}}_0 +\frac{v}{2}\int d\mbox{\boldmath$r$}\left\{\delta \hat{\phi}(\mbox{\boldmath$r$}) \right\}^2+ \frac{V}{2} \int d\mbox{\boldmath$r$}\hat{S}_{\alpha\beta}(\mbox{\boldmath$r$})\hat{S}_{\alpha\beta}(\mbox{\boldmath$r$})\\
&=\hat{\mathcal{H}}_0+\frac{v}{2}\sum_{j,k}\int d\mbox{\boldmath$r$}
\delta(\mbox{\boldmath$r$}-\mbox{\boldmath$r$}^{(j)})\delta(\mbox{\boldmath$r$}-\mbox{\boldmath$r$}^{(k)})+\frac{V}{2}\sum_{j,k}\int d\mbox{\boldmath$r$}\hat{s}_{\alpha\beta}^{(j)}\hat{s}_{\alpha\beta}^{(k)}
\delta(\mbox{\boldmath$r$}-\mbox{\boldmath$r$}^{(j)})\delta(\mbox{\boldmath$r$}-\mbox{\boldmath$r$}^{(k)})\\
&=\hat{\mathcal{H}}_0+\hat{\mathcal{W}}_\phi+\hat{\mathcal{W}}_S,  
\end{align*}
where
\begin{align*}
\hat{\mathcal{W}}_\phi&=\frac{v}{2}\int d\mbox{\boldmath$r$}\left\{\delta \hat{\phi}(\mbox{\boldmath$r$}) \right\}^2\\
\hat{\mathcal{W}}_S&=\frac{V}{2} \int d\mbox{\boldmath$r$}\hat{S}_{\alpha\beta}(\mbox{\boldmath$r$})\hat{S}_{\alpha\beta}(\mbox{\boldmath$r$}).
\end{align*}
The partition function $Z$ of the total system is written as
\begin{align}
Z&=\int d\Gamma \exp{\left(-\beta \hat{\mathcal{H}}\right)}\nonumber\\
&=\int d\mbox{\boldmath$r$}^{(0)}d\mbox{\boldmath$b$}^{(0)} \cdots d\mbox{\boldmath$b$}^{(N)}\exp{\left(-\beta \hat{\mathcal{H}}\right)},\label{eqn:partition_function}
\end{align}
where $\Gamma=\left\{\mbox{\boldmath$r$}^{(0)},\mbox{\boldmath$b$}^{(0)}, \mbox{\boldmath$b$}^{(1)} \cdots \mbox{\boldmath$b$}^{(N)}  \right\}$ is the point in the phase space, $\mbox{\boldmath$r$}^{(0)}$ is the position of the first bond, and $\mbox{\boldmath$b$}^{(i)}$ is the $i$-th bond vector.
In our model, $e^{-\beta \hat{\mathcal{H}}_0}$ corresponds to the transfer matrix $\mathcal{T}$.
We rewrite eq.(\ref{eqn:partition_function}) as follows;
\begin{align}
Z&=\int d\mbox{\boldmath$b$}^{(0)} \cdots d\mbox{\boldmath$b$}^{(N)}\exp{\left(-\beta \hat{\mathcal{H}}\right)}\nonumber\\
&=\int d\mbox{\boldmath$b$}^{(0)} \cdots d\mbox{\boldmath$b$}^{(N)} d\mbox{\boldmath$r$}^{(0)}\cdots d\mbox{\boldmath$r$}^{(N)} \prod_{i=0}^{N-1} \delta\left(\mbox{\boldmath$r$}^{(i+1)}-\mbox{\boldmath$r$}^{(i)}-\left(\frac{1}{2}\mbox{\boldmath$b$}^{(i)}+\frac{1}{2}\mbox{\boldmath$b$}^{(i+1)}  \right)  \right)\exp{\left(-\beta\hat{\mathcal{H}}  \right)}\label{eqn:partition_function_with_delta}.
\end{align}
In this equation, we can regard the orientation and the position of a segment as independent variables.
Then, we can obtain the effective Hamiltonian $\hat{\mathcal{H}}^{(\rm eff)}$ by eliminating the degrees of freedom of the orientation in eq.~(\ref{eqn:partition_function_with_delta}).
When the interaction energies $\hat{\mathcal{W}}_\phi$ and $\hat{\mathcal{W}}_S$ are small compared to the thermal energy $k_{\rm B}T$, we can approximate the Boltzman factor as
\begin{align}
\exp{\left[-\beta\left( \hat{\mathcal{H}}_0+\hat{\mathcal{W}}_\phi+\hat{\mathcal{W}}_S \right)\right]}&=\exp{\left(-\beta \hat{\mathcal{H}}_0\right)}\left(1-\beta \hat{\mathcal{W}}_\phi-\beta \hat{\mathcal{W}}_S  \right).
\end{align}
Thus, we should consider 
\begin{align}
&\int  d\mbox{\boldmath$r$}^{(0)}\cdots d\mbox{\boldmath$r$}^{(N)}\int d\mbox{\boldmath$b$}^{(0)} \cdots d\mbox{\boldmath$b$}^{(N)} \prod_{i=0}^{N-1} \delta\left(\mbox{\boldmath$r$}^{(i+1)}-\mbox{\boldmath$r$}^{(i)}-\left(\frac{1}{2}\mbox{\boldmath$b$}^{(i)}+\frac{1}{2}\mbox{\boldmath$b$}^{(i+1)}  \right)  \right)\exp{\left(-\beta\hat{\mathcal{H}}_0  \right)}\left(1-\beta \hat{\mathcal{W}}_S  \right)\nonumber\\
&=\int  d\mbox{\boldmath$r$}^{(0)}\cdots d\mbox{\boldmath$r$}^{(N)}Z_0^\prime \left\langle \left(1-\beta \hat{\mathcal{W}}_S \right) \right\rangle_0^\prime \label{eqn:compare1}\\
&=\int  d\mbox{\boldmath$r$}^{(0)}\cdots d\mbox{\boldmath$r$}^{(N)}\exp{\left(-\beta \hat{\mathcal{H}}_0^{\rm (eff)} \right)}\left(1-\beta \hat{\mathcal{W}}_{\phi}^{(\rm eff)}  \right)\label{eqn:compare2},
\end{align}
where $\langle \cdots \rangle_0^\prime$ means the average over only orientation in the reference state.
By comparing eqs.~(\ref{eqn:compare1}) and (\ref{eqn:compare2}), we obtain the relationship
\begin{align}
\beta\exp{\left(-\beta \hat{\mathcal{H}}_0^\prime \right)}\hat{\mathcal{W}}_{\phi}^{\rm (eff)}&=\beta Z_0^\prime \langle \hat{\mathcal{W}}_S \rangle_0^\prime,
\end{align}
where $Z_0^\prime$ is the partial summation of $\exp{\left[- \beta \hat{\mathcal{H}}_0\right]}$ over only the orientation.
Right hand side of this equation is  
\begin{align}
&\beta Z_0^\prime \langle \hat{\mathcal{W}}_S  \rangle_0^\prime \nonumber \\
&=\beta V\sum_{j,k=0}^{N}\int d\mbox{\boldmath$b$}^{(0)}d\mbox{\boldmath$b$}^{(1)}\cdots d\mbox{\boldmath$b$}^{(N)}\hat{s}_{\alpha\beta}^{(j)}
\hat{s}_{\alpha\beta}^{(k)}\delta(\mbox{\boldmath$r$}-\mbox{\boldmath$r$}^{(j)})
\delta(\mbox{\boldmath$r$}-\mbox{\boldmath$r$}^{(k)})e^{-\beta \hat{\mathcal{H}}_0}\prod_{i=0}^{N-1} \delta\left(\mbox{\boldmath$r$}^{(i+1)}-\mbox{\boldmath$r$}^{(i)}-\left(\frac{1}{2}\mbox{\boldmath$b$}^{(i)}+\frac{1}{2}\mbox{\boldmath$b$}^{(i+1)}  \right)  \right)\nonumber\\
&=\beta V\sum_{j,k=0}^{N}\int d\mbox{\boldmath$b$}^{(0)}d\mbox{\boldmath$b$}^{(1)}\cdots d\mbox{\boldmath$b$}^{(N)} \prod_{i=0}^{N-1}\sum_{\mu^{(i)}}Q(i,\mu^{(i)},\mbox{\boldmath$r$}^{(i)};i+1,\mu^{(i+1)},\mbox{\boldmath$r$}^{(i+1)})
\hat{s}_{\alpha\beta}^{(j)}
\hat{s}_{\alpha\beta}^{(k)}\delta(\mbox{\boldmath$r$}-\mbox{\boldmath$r$}^{(j)})
\delta(\mbox{\boldmath$r$}-\mbox{\boldmath$r$}^{(k)})\label{eqn:ZW_S},
\end{align}
where $Q(i,\mu^{(i)},\mbox{\boldmath$r$}^{(i)};i+1,\mu^{(i+1)},\mbox{\boldmath$r$}^{(i+1)})$ is the path integral whose variables are the positions and the orientations of the bonds:
\begin{align}
Q(i,\mu^{(i)},\mbox{\boldmath$r$}^{(i)};i+1,\mu^{(i+1)},\mbox{\boldmath$r$}^{(i+1)})&=\mathcal{T}_{\mu^{(i)}\mu^{(i+1)}}
\delta\left(\mbox{\boldmath$r$}^{(i+1)}-\mbox{\boldmath$r$}^{(i)}
-\frac{\mbox{\boldmath$b$}^{(i)}+\mbox{\boldmath$b$}^{(i+1)}}{2}  \right)\nonumber\\
&=\int d\mbox{\boldmath$q$} \tilde{\mathcal{T}}_{\mu^{(i)}\mu^{(i+1)}}(\mbox{\boldmath$q$})\exp{\left[i\mbox{\boldmath$q$}\cdot (\mbox{\boldmath$r$}^{(i+1)}-\mbox{\boldmath$r$}^{(i)})  \right]}\\
\tilde{\mathcal{T}}_{\mu\nu}(\mbox{\boldmath$q$})&=\exp{\left[-i\mbox{\boldmath$q$}\cdot\frac{\mbox{\boldmath$e$}_\mu}{2}\right]}T_{\mu\nu}\exp{\left[-i\mbox{\boldmath$q$}\cdot\frac{\mbox{\boldmath$e$}_\nu}{2}\right]}.
\end{align}
By using this path integral, we can rewrite $\left\langle \hat{\mathcal{W}}_S \right\rangle_0^\prime$ in terms of the transfer matrix in the Fourier space as 
\begin{align}
&Z_0^\prime \langle \hat{\mathcal{W}}_S  \rangle_0^\prime\nonumber\\
&=V\sum_{j,k=0}^{N}\int d\mbox{\boldmath$b$}^{(0)}d\mbox{\boldmath$b$}^{(1)}\cdots d\mbox{\boldmath$b$}^{(N)} \prod_{i=0}^{N-1}\sum_{\mu^{(i)}}Q(i,\mu^{(i)},\mbox{\boldmath$r$}^{(i)};i+1,\mu^{(i+1)},\mbox{\boldmath$r$}^{(i+1)})
\hat{s}_{\alpha\beta}^{(j)}
\hat{s}_{\alpha\beta}^{(k)}\delta(\mbox{\boldmath$r$}-\mbox{\boldmath$r$}^{(j)})
\delta(\mbox{\boldmath$r$}-\mbox{\boldmath$r$}^{(k)})\nonumber\\
&=V\sum_{j,k=0}^{N}\int d\mbox{\boldmath$b$}^{(0)}d\mbox{\boldmath$b$}^{(1)}\cdots d\mbox{\boldmath$b$}^{(N)} \prod_{i=0}^{N-1}\sum_{\mu^{(i)}}\int d\mbox{\boldmath$q$}_i \tilde{\mathcal{T}}_{\mu^{(i)}\mu^{(i+1)}}({\mbox{\boldmath$q$}_i})
\exp{\left[ i\mbox{\boldmath$q$}_i\cdot\left(\mbox{\boldmath$r$}^{(i+1)}-\mbox{\boldmath$r$}^{(i)} \right)\right]}
\hat{s}_{\alpha\beta}^{(j)}
\hat{s}_{\alpha\beta}^{(k)}\nonumber\\
&\times \delta(\mbox{\boldmath$r$}-\mbox{\boldmath$r$}^{(j)})
\delta(\mbox{\boldmath$r$}-\mbox{\boldmath$r$}^{(k)})\nonumber\\
&=V\prod_{i=0}^{N-1}\sum_{\mu^(i)}\sum_{j,k=0}^{N}\int d\mbox{\boldmath$b$}^{(0)}\cdots d\mbox{\boldmath$b$}^{(N)}
\tilde{\mathcal{T}}_{\mu^{(i)}\mu^{(i+1)}}({\mbox{\boldmath$q$}_i})\hat{s}_{\alpha\beta}^{(j)}\hat{s}_{\alpha\beta}^{(k)}\exp{\left[ i\mbox{\boldmath$q$}_i\cdot\left(\mbox{\boldmath$r$}^{(i+1)}-\mbox{\boldmath$r$}^{(i)} \right)\right]}\nonumber\\
&\times \delta(\mbox{\boldmath$r$}-\mbox{\boldmath$r$}^{(j)})
\delta(\mbox{\boldmath$r$}-\mbox{\boldmath$r$}^{(k)})\nonumber\\
&=V\sum_{j,k=0}^{N}\int d\mbox{\boldmath$b$}^{(0)}\cdots d\mbox{\boldmath$b$}^{(N)} \tilde{\mathcal{Q}}(\Gamma;\left\{\mbox{\boldmath$q$}_i \right\})\exp{\left[ i\mbox{\boldmath$q$}_i\cdot\left(\mbox{\boldmath$r$}^{(i+1)}-\mbox{\boldmath$r$}^{(i)} \right)\right]}\delta(\mbox{\boldmath$r$}-\mbox{\boldmath$r$}^{(j)})
\delta(\mbox{\boldmath$r$}-\mbox{\boldmath$r$}^{(k)}),
\end{align}
where
\begin{align}
\tilde{\mathcal{Q}}(\Gamma;\left\{\mbox{\boldmath$q$}_i \right\})&=\prod_{i=0}^{N-1}\sum_{\mu^(i)}\tilde{\mathcal{T}}_{\mu^{(i)}\mu^{(i+1)}}(\mbox{\boldmath$q$}_i)
\end{align}
We should consider how the {\it nematic} {\it interaction} affects the {\it excluded volume} {\it interaction}.
As the self terms($j=k$) give constant contribution to the final result, these terms can be dropped.
When we focus on the statistical weight $\tilde{\mathcal{Q}}(\Gamma; \left\{\mbox{\boldmath$q$} \right\}) \propto \exp{\left[i\mbox{\boldmath$q$}_i \cdot \left(\frac{\mbox{\boldmath$b$}^{(i+1)}+\mbox{\boldmath$b$}^{(i)}}{\displaystyle{2}} \right) \right]}$ which plays the role of the moment generating function for the orientations, we can obtain the correlation function by considering the symmetry of the system.
By using this equation and symmetry of the system, we can obtain 
\begin{align}
&\left(i\frac{\partial}{\partial \mbox{\boldmath$q$}_j}\right)\left(i\frac{\partial}{\partial \mbox{\boldmath$q$}_j}\right)\left(i\frac{\partial}{\partial \mbox{\boldmath$q$}_k}\right)\left(i\frac{\partial}{\partial \mbox{\boldmath$q$}_k}\right)\tilde{Z}_0^\prime(\left\{ \mbox{\boldmath$q$}_i \right\}) \nonumber\\
&=\left(i\frac{\partial}{\partial \mbox{\boldmath$q$}_j}\right)\left(i\frac{\partial}{\partial \mbox{\boldmath$q$}_j}\right)\left(i\frac{\partial}{\partial \mbox{\boldmath$q$}_k}\right)\left(i\frac{\partial}{\partial \mbox{\boldmath$q$}_k}\right)\int d\mbox{\boldmath$b$}^{(0)}\cdots d\mbox{\boldmath$b$}^{(N)} \tilde{\mathcal{Q}}(\Gamma;\left\{ \mbox{\boldmath$q$}_i \right\})\nonumber\\
&=\frac{1}{16}\prod_{i=0}^{N-1} \sum_{\mu^{(i)}}\int d\mbox{\boldmath$b$}^{(0)}\cdots d\mbox{\boldmath$b$}^{(N)} \tilde{\mathcal{Q}}(\Gamma;\left\{ \mbox{\boldmath$q$}_i \right\})\nonumber\\
&\left(\mbox{\boldmath$b$}^{(j)}\mbox{\boldmath$b$}^{(j)}\mbox{\boldmath$b$}^{(k)}\mbox{\boldmath$b$}^{(k)}
+\mbox{\boldmath$b$}^{(j)}\mbox{\boldmath$b$}^{(j)}\mbox{\boldmath$b$}^{(k+1)}\mbox{\boldmath$b$}^{(k+1)}\right.\nonumber\\
&\left. +\mbox{\boldmath$b$}^{(j+1)}\mbox{\boldmath$b$}^{(j+1)}\mbox{\boldmath$b$}^{(k)}\mbox{\boldmath$b$}^{(k)}
+\mbox{\boldmath$b$}^{(j+1)}\mbox{\boldmath$b$}^{(j+1)}\mbox{\boldmath$b$}^{(k+1)}\mbox{\boldmath$b$}^{(k+1)}
\right)\nonumber\\
&=\frac{1}{16}\left\{\prod_{i=0}^{N-1} \sum_{\mu^{(i)}}\int d\mbox{\boldmath$b$}^{(0)}\cdots d\mbox{\boldmath$b$}^{(N)} \tilde{\mathcal{Q}}(\Gamma;\left\{ \mbox{\boldmath$q$}_i \right\})\right.\nonumber\\
&\left.\times \left( 2+2\cosh{\left[ \frac{3}{l_{\rm p}} \right]} \right)\mbox{\boldmath$b$}^{(j)}\mbox{\boldmath$b$}^{(j)}\mbox{\boldmath$b$}^{(k)}\mbox{\boldmath$b$}^{(k)}\right\}\label{eqn:moment_generating},
\end{align}
where we use $\left\langle \left(\mbox{\boldmath$b$}^{(j)}\cdot\mbox{\boldmath$b$}^{(k)}  \right)^2 \right\rangle\sim \exp{\left[3|j-k|/l_{\rm p}\right]}$ which is determined by the fitting.
It should be noted that this value is different from the result of the Gaussian distribution $\left\langle \left(\mbox{\boldmath$b$}^{(j)}\cdot\mbox{\boldmath$b$}^{(k)}  \right)^2 \right\rangle\sim \exp{\left[2|j-k|/l_{\rm p}\right]}$. 
The difference between Gaussian and our model does not affect the final result qualitatively because their behaviors in the small wave number region, which is essential to the results of the present study, are qualitatively the same. 
By using eq.~(\ref{eqn:moment_generating}), four-body correlation of bonds satisfies  
\begin{align}
&\prod_{i=0}^{N-1} \sum_{\mu^{(i)}}\int d\mbox{\boldmath$b$}^{(0)}\cdots d\mbox{\boldmath$b$}^{(N)} \tilde{\mathcal{Q}}(\Gamma;\left\{ \mbox{\boldmath$q$}_i \right\})
\mbox{\boldmath$b$}^{(j)}\mbox{\boldmath$b$}^{(j)}\mbox{\boldmath$b$}^{(k)}\mbox{\boldmath$b$}^{(k)}\nonumber\\
&=\frac{8}{\displaystyle{1+\cosh{\left[ \frac{3}{l_{\rm p}} \right]}}}\times \left(i\frac{\partial}{\partial \mbox{\boldmath$q$}_j}\right)\left(i\frac{\partial}{\partial \mbox{\boldmath$q$}_j}\right)\left(i\frac{\partial}{\partial \mbox{\boldmath$q$}_k}\right)\left(i\frac{\partial}{\partial \mbox{\boldmath$q$}_k}\right)
\prod_{i=0}^{N-1} \sum_{\mu^{(i)}}\int d\mbox{\boldmath$b$}^{(0)}\cdots d\mbox{\boldmath$b$}^{(N)} \tilde{\mathcal{Q}}(\Gamma;\left\{ \mbox{\boldmath$q$}_i \right\})\label{eqn:bbbb}.
\end{align}
The orientation order parameter can be written in terms of the bond vectors as follows:
\begin{align}
\hat{s}_{\alpha\beta}^{(j)}\hat{s}_{\alpha\beta}^{(k)}&=\frac{9}{4}\left(b_\alpha^{(j)} b_\beta^{(j)}-\frac{\delta_{\alpha\beta}}{3}  \right)\left(b_\alpha^{(k)} b_\beta^{(k)}-\frac{\delta_{\alpha\beta}}{3}  \right)\nonumber\\
&=\frac{9}{4}\left(b_\alpha^{(j)} b_\beta^{(j)}b_\alpha^{(k)} b_\beta^{(k)}-\frac{b_{\alpha}^{(j)}b_{\alpha}^{(j)}}{3}- \frac{b_{\alpha}^{(k)}b_{\alpha}^{(k)}}{3}+\frac{1}{3} \right)\nonumber\\
&=\frac{9}{4}\left(b_\alpha^{(j)} b_\beta^{(j)}b_\alpha^{(k)} b_\beta^{(k)}-\frac{1}{3} \right).
\end{align}
Then, we obtain 
\begin{align}
&\prod_{i=0}^{N-1} \sum_{\mu^{(i)}}\int d\mbox{\boldmath$b$}^{(0)}\cdots d\mbox{\boldmath$b$}^{(N)} \tilde{\mathcal{Q}}(\Gamma;\left\{ \mbox{\boldmath$q$}_i \right\})
\hat{s}_{\alpha\beta}^{(j)}\hat{s}_{\alpha\beta}^{(k)}\nonumber \\
&=\frac{9}{4}\left({\rm The\ components\ of\ }\alpha\beta\alpha\beta{\ {\rm in}\ {\rm eq.~}(\ref{eqn:bbbb})}  \right)-\frac{3}{4}\prod_{i=0}^{N-1} \sum_{\mu^{(i)}}\int d\mbox{\boldmath$b$}^{(0)}\cdots d\mbox{\boldmath$b$}^{(N)} \tilde{\mathcal{Q}}(\Gamma;\left\{ \mbox{\boldmath$q$}_i \right\})\label{eqn:ss}.
\end{align}
By using eqs.~(\ref{eqn:ZW_S}) and (\ref{eqn:bbbb})-(\ref{eqn:ss}), we evaluate eq.(\ref{eqn:compare2}) and compare the result of this evaluation with eq.~(\ref{eqn:compare1}), we obtain
\begin{align}
&\int d\mbox{\boldmath$r$}^{(0)}\cdots d\mbox{\boldmath$r$}^{(N)}\beta Z_0^\prime \langle \hat{\mathcal{W}}_S\rangle^\prime\nonumber\\
&=
\beta V\sum_{j,k}\int d\mbox{\boldmath$r$}^{(0)}\cdots d\mbox{\boldmath$r$}^{(N)}\left\{\frac{9}{4}\frac{8}{\displaystyle{1+\cosh{\left[\frac{3}{l_{\rm p}} \right]}}} \left\{\left(\mbox{\boldmath$r$}^{(j+1)}-\mbox{\boldmath$r$}^{(j)} \right)\cdot \left(\mbox{\boldmath$r$}^{(k+1)}-\mbox{\boldmath$r$}^{(k)} \right)\right\}^2-\frac{3}{4} \right\} Z_0^\prime\nonumber \\
&\times \delta\left(\mbox{\boldmath$r$}-\mbox{\boldmath$r$}^{(j)} \right)\delta\left(\mbox{\boldmath$r$}-\mbox{\boldmath$r$}^{(k)} \right)\nonumber\\
&=V\sum_{j,k}\int d\mbox{\boldmath$r$}^{(0)}\cdots d\mbox{\boldmath$r$}^{(N)}
\exp{\left[ -\beta \hat{\mathcal{H}}_0^\prime\right]}\nonumber\\
&\left\{\frac{9}{4}\frac{8}{\displaystyle{1+\cosh{\left[\frac{3}{l_{\rm p}} \right]}}} \left\{\left(\mbox{\boldmath$r$}^{(j+1)}-\mbox{\boldmath$r$}^{(j)} \right)\cdot \left(\mbox{\boldmath$r$}^{(k+1)}-\mbox{\boldmath$r$}^{(k)} \right)\right\}^2-\frac{3}{4} \right\} \nonumber \\
&\times \delta\left(\mbox{\boldmath$r$}-\mbox{\boldmath$r$}^{(j)} \right)\delta\left(\mbox{\boldmath$r$}-\mbox{\boldmath$r$}^{(k)} \right)\nonumber\\
&=\int d\mbox{\boldmath$r$}^{(0)}\cdots d\mbox{\boldmath$r$}^{(N)} \beta \exp{\left[-\beta \hat{\mathcal{H}}_0^\prime\right]}\hat{\mathcal{W}}_\phi^{\rm (ne)},
\end{align}
where 
\begin{align}
\exp{\left[-\beta \hat{\mathcal{H}}_0^\prime  \right]}&=\exp{\left[\ln{\left[Z_0^\prime \right]}\right]}\\
\hat{\mathcal{W}}_\phi^{\rm (ne)}&= V\int d\mbox{\boldmath$r$} \sum_{j\neq k}\left\{\frac{18}{\displaystyle{1+\cosh{\left[\frac{3}{l_{\rm p}} \right]}}} \left\{\left(\mbox{\boldmath$r$}^{(j+1)}-\mbox{\boldmath$r$}^{(j)} \right)\cdot \left(\mbox{\boldmath$r$}^{(k+1)}-\mbox{\boldmath$r$}^{(k)} \right)\right\}^2-\frac{3}{4} \right\} \nonumber \\
&\times \delta\left(\mbox{\boldmath$r$}-\mbox{\boldmath$r$}^{(j)} \right)\delta\left(\mbox{\boldmath$r$}-\mbox{\boldmath$r$}^{(k)} \right)\label{eqn:nematic_in_exvol}.
\end{align}
The contact interaction derived from the nematic interaction depends on the stiffness of the polymer chain.
In the next appendix, we estimate these terms by using preaveraging approximation.

\subsection{Preaveraging approximation}\label{sec:preave}
Our starting point is the Hamiltonian
\begin{align}
\hat{\mathcal{H}}^{\rm (eff)}=\hat{\mathcal{H}}_0^{(\rm eff)}+\hat{\mathcal{W}}_\phi^{\prime}, 
\end{align}
where $\hat{\mathcal{W}}_\phi^{\prime}$ includes the excluded volume and the effect of the nematic interaction eq.~(\ref{eqn:nematic_in_exvol}) given by
\begin{align}
W_\phi^{\prime}&=\frac{v}{2}\int d\mbox{\boldmath$r$} \left\{ \delta\hat{\phi}(\mbox{\boldmath$r$})\right\}^2\nonumber\\
&+ V\int d\mbox{\boldmath$r$}\sum_{j< k}\left\{\frac{18}{\displaystyle{1+\cosh{\left[\frac{3}{l_{\rm p}} \right]}}} \left\{\left(\mbox{\boldmath$r$}^{(j+1)}-\mbox{\boldmath$r$}^{(j)} \right)\cdot \left(\mbox{\boldmath$r$}^{(k+1)}-\mbox{\boldmath$r$}^{(k)} \right)\right\}^2-\frac{3}{4} \right\} \delta\left(\mbox{\boldmath$r$}-\mbox{\boldmath$r$}^{(j)} \right)\delta\left(\mbox{\boldmath$r$}-\mbox{\boldmath$r$}^{(k)} \right).\label{eqn:renormalized_interaction}
\end{align}
We apply the preaveraging approximation which replaces $\left\{ \left(\mbox{\boldmath$r$}^{(j+1)}-\mbox{\boldmath$r$}^{(j)} \right)\cdot \left(\mbox{\boldmath$r$}^{(k+1)}-\mbox{\boldmath$r$}^{(k)} \right) \right\}^2$  as
\begin{align}
\left\{ \left(\mbox{\boldmath$r$}^{(j+1)}-\mbox{\boldmath$r$}^{(j)} \right)\cdot \left(\mbox{\boldmath$r$}^{(k+1)}-\mbox{\boldmath$r$}^{(k)} \right) \right\}^2&\rightarrow
\left\langle\left\{\left(\mbox{\boldmath$r$}^{(j+1)}-\mbox{\boldmath$r$}^{(j)} \right)\cdot \left(\mbox{\boldmath$r$}^{(k+1)}-\mbox{\boldmath$r$}^{(k)} \right)\right\}^2 \right\rangle\nonumber\\
&=\frac{1}{16}\left\langle\left\{\left(\mbox{\boldmath$b$}^{(j)}+\mbox{\boldmath$b$}^{(j+1)} \right)\cdot \left(\mbox{\boldmath$b$}^{(k)}-\mbox{\boldmath$b$}^{(k+1)} \right)\right\}^2 \right\rangle\nonumber\\
&=\frac{1}{16} \left\{\left\langle\left(\mbox{\boldmath$b$}^{(j)}\cdot\mbox{\boldmath$b$}^{(k)}\right)^2\right\rangle
+\left\langle\left(\mbox{\boldmath$b$}^{(j+1)}\cdot\mbox{\boldmath$b$}^{(k)}\right)^2\right\rangle\right.\nonumber\\
&\left.+\left\langle\left(\mbox{\boldmath$b$}^{(j)}\cdot\mbox{\boldmath$b$}^{(k+1)}\right)^2\right\rangle
+\left\langle\left(\mbox{\boldmath$b$}^{(j+1)}\cdot\mbox{\boldmath$b$}^{(k+1)}\right)^2\right\rangle \right\}\label{eqn:preaveraging},
\end{align}
where the coupling terms are dropped by considering the symmetry of the system.
The expression eq.~(\ref{eqn:preaveraging}) is written as
\begin{align}
&\frac{1}{16} \left\{\left\langle\left(\mbox{\boldmath$b$}^{(j)}\cdot\mbox{\boldmath$b$}^{(k)}\right)^2\right\rangle
+\left\langle\left(\mbox{\boldmath$b$}^{(j+1)}\cdot\mbox{\boldmath$b$}^{(k)}\right)^2\right\rangle+\left\langle\left(\mbox{\boldmath$b$}^{(j)}\cdot\mbox{\boldmath$b$}^{(k+1)}\right)^2\right\rangle
+\left\langle\left(\mbox{\boldmath$b$}^{(j+1)}\cdot\mbox{\boldmath$b$}^{(k+1)}\right)^2\right\rangle\right\}\nonumber\\
&=\frac{1}{16}\left\{\exp{\left[-\frac{3|k-j  |}{l_{\rm p}}\right]}
+\exp{\left[-\frac{3|k-j-1|}{l_{\rm p}}\right]}+\exp{\left[-\frac{3|k+1-j|}{l_{\rm p}}\right]}
+\exp{\left[-\frac{3|k-j|}{l_{\rm p}}\right]}\right\}\nonumber\\
&=\frac{1}{8}\exp{\left[-3\frac{k-j}{l_{\rm p}}  \right]}\left\{1+\cosh{\left[\frac{3}{l_{\rm p}} \right]} \right\}.
\end{align}
Therefore, we obtain
\begin{align}
\hat{\mathcal{W}}_\phi^{\prime}&=\frac{v}{2}\int d\mbox{\boldmath$r$} \left\{ \delta\hat{\phi}(\mbox{\boldmath$r$})\right\}^2\nonumber\\
&+ V\int d\mbox{\boldmath$r$}\sum_{j< k}\left\{\frac{18}{\displaystyle{1+\cosh{\left[\frac{3}{l_{\rm p}} \right]}}}\left( \frac{1}{8}\exp{\left[-3\frac{k-j}{l_{\rm p}}  \right]}\right)\left(1+\cosh{\left[\frac{3}{l_{\rm p}} \right]}\right)-\frac{3}{4}  \right\}\nonumber\\
&\times \delta\left(\mbox{\boldmath$r$}-\mbox{\boldmath$r$}^{(j)} \right)\delta\left(\mbox{\boldmath$r$}-\mbox{\boldmath$r$}^{(k)} \right)\nonumber\\
&=\frac{v}{2}\int d\mbox{\boldmath$r$} \left\{ \delta\hat{\phi}(\mbox{\boldmath$r$})\right\}^2\nonumber\\
&+ V\int d\mbox{\boldmath$r$}\sum_{j< k}\left\{\frac{9}{4}\exp{\left[-3\frac{k-j}{l_{\rm p}}  \right]}-\frac{3}{4}\right\} \delta\left(\mbox{\boldmath$r$}-\mbox{\boldmath$r$}^{(j)} \right)\delta\left(\mbox{\boldmath$r$}-\mbox{\boldmath$r$}^{(k)} \right)\label{eqn:after_preaverage}.
\end{align}
When we assume the uniform distribution for the positions of the segments, we estimate eq.~(\ref{eqn:after_preaverage}) as
\begin{align}
\hat{\mathcal{W}}_\phi^{\prime}&=\frac{v}{2}\int d\mbox{\boldmath$r$} \left\{ \delta\hat{\phi}(\mbox{\boldmath$r$})\right\}^2\nonumber\\
&+ V\times N\frac{1}{N^2}\sum_{j< k}\left\{\frac{9}{4}\exp{\left[-3\frac{k-j}{l_{\rm p}}\right]}-\frac{3}{4}\right\}\int d\mbox{\boldmath$r$} \delta\left(\mbox{\boldmath$r$}-\mbox{\boldmath$r$}^{(j)} \right)\delta\left(\mbox{\boldmath$r$}-\mbox{\boldmath$r$}^{(k)} \right)\nonumber\\
&=\frac{v}{2}\int d\mbox{\boldmath$r$} \left\{ \hat{\phi}(\mbox{\boldmath$r$})\right\}^2\nonumber\\
&+ \underline{\left[\frac{V}{N}\sum_{j< k}\left\{\frac{9}{4}\exp{\left[-3\frac{k-j}{l_{\rm p}}  \right]}\right\}-\frac{3}{4}\right]}\int d\mbox{\boldmath$r$} \left\{\delta \hat{\phi}(\mbox{\boldmath$r$}) \right\}^2.
\end{align}
The underlined part describes the effect of the nematic interaction on the excluded volume interaction.

\section*{Acknowledgement}
The authors are indebted to Professor Imai for valuable discussions. 
This work is partially supported by the Grant-in-Aid for Scientific Research (Grant Number 26287096) from The Ministry of Education, Culture, Sports, Science and Technology(MEXT), Japan.

\bibliographystyle{achemso}
\bibliography{ref}

\providecommand{\latin}[1]{#1}
\providecommand*\mcitethebibliography{\thebibliography}
\csname @ifundefined\endcsname{endmcitethebibliography}
  {\let\endmcitethebibliography\endthebibliography}{}
\begin{mcitethebibliography}{0}
\providecommand*\natexlab[1]{#1}
\providecommand*\mciteSetBstSublistMode[1]{}
\providecommand*\mciteSetBstMaxWidthForm[2]{}
\providecommand*\mciteBstWouldAddEndPuncttrue
  {\def\EndOfBibitem{\unskip.}}
\providecommand*\mciteBstWouldAddEndPunctfalse
  {\let\EndOfBibitem\relax}
\providecommand*\mciteSetBstMidEndSepPunct[3]{}
\providecommand*\mciteSetBstSublistLabelBeginEnd[3]{}
\providecommand*\EndOfBibitem{}
\mciteSetBstSublistMode{f}
\mciteSetBstMaxWidthForm{subitem}{(\alph{mcitesubitemcount})}
\mciteSetBstSublistLabelBeginEnd
  {\mcitemaxwidthsubitemform\space}
  {\relax}
  {\relax}

\end{mcitethebibliography}


\providecommand{\latin}[1]{#1}
\providecommand*\mcitethebibliography{\thebibliography}
\csname @ifundefined\endcsname{endmcitethebibliography}
  {\let\endmcitethebibliography\endthebibliography}{}
\begin{mcitethebibliography}{25}
\providecommand*\natexlab[1]{#1}
\providecommand*\mciteSetBstSublistMode[1]{}
\providecommand*\mciteSetBstMaxWidthForm[2]{}
\providecommand*\mciteBstWouldAddEndPuncttrue
  {\def\EndOfBibitem{\unskip.}}
\providecommand*\mciteBstWouldAddEndPunctfalse
  {\let\EndOfBibitem\relax}
\providecommand*\mciteSetBstMidEndSepPunct[3]{}
\providecommand*\mciteSetBstSublistLabelBeginEnd[3]{}
\providecommand*\EndOfBibitem{}
\mciteSetBstSublistMode{f}
\mciteSetBstMaxWidthForm{subitem}{(\alph{mcitesubitemcount})}
\mciteSetBstSublistLabelBeginEnd
  {\mcitemaxwidthsubitemform\space}
  {\relax}
  {\relax}

\bibitem[Alder and Wainwright(1957)Alder, and Wainwright]{Alder}
Alder,~B.~J.; Wainwright,~T.~E. \emph{J. Chem. Phys.}
  \textbf{1957}, \emph{27}, 1208--1209\relax
\mciteBstWouldAddEndPuncttrue
\mciteSetBstMidEndSepPunct{\mcitedefaultmidpunct}
{\mcitedefaultendpunct}{\mcitedefaultseppunct}\relax
\EndOfBibitem
\bibitem[Yamamoto(2008)]{Yamamoto2008}
Yamamoto,~T. \emph{J.Chem.Phys.} \textbf{2008}, \emph{129},
  184903\relax
\mciteBstWouldAddEndPuncttrue
\mciteSetBstMidEndSepPunct{\mcitedefaultmidpunct}
{\mcitedefaultendpunct}{\mcitedefaultseppunct}\relax
\EndOfBibitem
\bibitem[Milner(2011)]{milner}
Milner,~S.~T. \emph{Soft Matter} \textbf{2011}, \emph{7}, 2909--2917\relax
\mciteBstWouldAddEndPuncttrue
\mciteSetBstMidEndSepPunct{\mcitedefaultmidpunct}
{\mcitedefaultendpunct}{\mcitedefaultseppunct}\relax
\EndOfBibitem
\bibitem[Xu \latin{et~al.}(2005)Xu, Matkar, and Kyu]{phase}
Xu,~H.; Matkar,~R.; Kyu,~T. \emph{Phys. Rev. E} \textbf{2005}, \emph{72},
  011804\relax
\mciteBstWouldAddEndPuncttrue
\mciteSetBstMidEndSepPunct{\mcitedefaultmidpunct}
{\mcitedefaultendpunct}{\mcitedefaultseppunct}\relax
\EndOfBibitem
\bibitem[Imai \latin{et~al.}(1992)Imai, Mori, Mizukami, Kaji, and
  Kanaya]{imai1}
Imai,~M.; Mori,~K.; Mizukami,~T.; Kaji,~K.; Kanaya,~T. \emph{Polymer}
  \textbf{1992}, \emph{33}, 4451 -- 4456\relax
\mciteBstWouldAddEndPuncttrue
\mciteSetBstMidEndSepPunct{\mcitedefaultmidpunct}
{\mcitedefaultendpunct}{\mcitedefaultseppunct}\relax
\EndOfBibitem
\bibitem[Imai \latin{et~al.}(1992)Imai, Mori, Mizukami, Kaji, and
  Kanaya]{imai2}
Imai,~M.; Mori,~K.; Mizukami,~T.; Kaji,~K.; Kanaya,~T. \emph{Polymer}
  \textbf{1992}, \emph{33}, 4457 -- 4462\relax
\mciteBstWouldAddEndPuncttrue
\mciteSetBstMidEndSepPunct{\mcitedefaultmidpunct}
{\mcitedefaultendpunct}{\mcitedefaultseppunct}\relax
\EndOfBibitem
\bibitem[Imai \latin{et~al.}(1994)Imai, Kaji, and Kanaya]{imai3}
Imai,~M.; Kaji,~K.; Kanaya,~T. \emph{Macromolecules} \textbf{1994}, \emph{27},
  7103--7108\relax
\mciteBstWouldAddEndPuncttrue
\mciteSetBstMidEndSepPunct{\mcitedefaultmidpunct}
{\mcitedefaultendpunct}{\mcitedefaultseppunct}\relax
\EndOfBibitem
\bibitem[Shimada \latin{et~al.}(1988)Shimada, Doi, and Okano]{Doi1}
Shimada,~T.; Doi,~M.; Okano,~K. \emph{J.Chem.Phys.}
  \textbf{1988}, \emph{88}\relax
\mciteBstWouldAddEndPuncttrue
\mciteSetBstMidEndSepPunct{\mcitedefaultmidpunct}
{\mcitedefaultendpunct}{\mcitedefaultseppunct}\relax
\EndOfBibitem
\bibitem[Doi \latin{et~al.}(1988)Doi, Shimada, and Okano]{Doi2}
Doi,~M.; Shimada,~T.; Okano,~K. \emph{J.Chem.Phys.}
  \textbf{1988}, \emph{88}\relax
\mciteBstWouldAddEndPuncttrue
\mciteSetBstMidEndSepPunct{\mcitedefaultmidpunct}
{\mcitedefaultendpunct}{\mcitedefaultseppunct}\relax
\EndOfBibitem
\bibitem[Shimada \latin{et~al.}(1988)Shimada, Doi, and Okano]{Doi3}
Shimada,~T.; Doi,~M.; Okano,~K. \emph{J.Chem.Phys.}
  \textbf{1988}, \emph{88}\relax
\mciteBstWouldAddEndPuncttrue
\mciteSetBstMidEndSepPunct{\mcitedefaultmidpunct}
{\mcitedefaultendpunct}{\mcitedefaultseppunct}\relax
\EndOfBibitem
\bibitem[Matsuba \latin{et~al.}(2002)Matsuba, Kaji, Kanaya, and
  Nishida]{Matsuda}
Matsuba,~G.; Kaji,~K.; Kanaya,~T.; Nishida,~K. \emph{Phys. Rev. E}
  \textbf{2002}, \emph{65}, 061801\relax
\mciteBstWouldAddEndPuncttrue
\mciteSetBstMidEndSepPunct{\mcitedefaultmidpunct}
{\mcitedefaultendpunct}{\mcitedefaultseppunct}\relax
\EndOfBibitem
\bibitem[Gee \latin{et~al.}(2006)Gee, Lacevic, and Fried]{Gee}
Gee,~R.~H.; Lacevic,~N.; Fried,~L.~E. \emph{Nature Materials} \textbf{2006},
  \emph{5}, 39 -- 43\relax
\mciteBstWouldAddEndPuncttrue
\mciteSetBstMidEndSepPunct{\mcitedefaultmidpunct}
{\mcitedefaultendpunct}{\mcitedefaultseppunct}\relax
\EndOfBibitem
\bibitem[Olmsted \latin{et~al.}(1998)Olmsted, Poon, McLeish, Terrill, and
  Ryan]{Olmsted}
Olmsted,~P.~D.; Poon,~W. C.~K.; McLeish,~T. C.~B.; Terrill,~N.~J.; Ryan,~A.~J.
  \emph{Phys. Rev. Lett.} \textbf{1998}, \emph{81}, 373--376\relax
\mciteBstWouldAddEndPuncttrue
\mciteSetBstMidEndSepPunct{\mcitedefaultmidpunct}
{\mcitedefaultendpunct}{\mcitedefaultseppunct}\relax
\EndOfBibitem
\bibitem[Tan \latin{et~al.}(2003)Tan, Miao, and Yan]{Tan}
Tan,~H.; Miao,~B.; Yan,~D. \emph{J.Chem.Phys.}
  \textbf{2003}, \emph{119}\relax
\mciteBstWouldAddEndPuncttrue
\mciteSetBstMidEndSepPunct{\mcitedefaultmidpunct}
{\mcitedefaultendpunct}{\mcitedefaultseppunct}\relax
\EndOfBibitem
\bibitem[Wang \latin{et~al.}(2000)Wang, Hsiao, Sirota, Agarwal, and
  Srinivas]{Z-G-Wang}
Wang,~Z.-G.; Hsiao,~B.~S.; Sirota,~E.~B.; Agarwal,~P.; Srinivas,~S.
  \emph{Macromolecules} \textbf{2000}, \emph{33}, 978--989\relax
\mciteBstWouldAddEndPuncttrue
\mciteSetBstMidEndSepPunct{\mcitedefaultmidpunct}
{\mcitedefaultendpunct}{\mcitedefaultseppunct}\relax
\EndOfBibitem
\bibitem[Cahn and Hilliard(1958)Cahn, and Hilliard]{Cahn-Hilliard}
Cahn,~J.~W.; Hilliard,~J.~E. \emph{J.Chem.Phys.}
  \textbf{1958}, \emph{28}, 258--267\relax
\mciteBstWouldAddEndPuncttrue
\mciteSetBstMidEndSepPunct{\mcitedefaultmidpunct}
{\mcitedefaultendpunct}{\mcitedefaultseppunct}\relax
\EndOfBibitem
\bibitem[Avrami(1939)]{Avrami}
Avrami,~M. \emph{J.Chem.Phys.} \textbf{1939}, \emph{7},
  1103--1112\relax
\mciteBstWouldAddEndPuncttrue
\mciteSetBstMidEndSepPunct{\mcitedefaultmidpunct}
{\mcitedefaultendpunct}{\mcitedefaultseppunct}\relax
\EndOfBibitem
\bibitem[Panine \latin{et~al.}(2008)Panine, Cola, Sztucki, and
  Narayanan]{Panine}
Panine,~P.; Cola,~E.~D.; Sztucki,~M.; Narayanan,~T. \emph{Polymer}
  \textbf{2008}, \emph{49}, 676 -- 680\relax
\mciteBstWouldAddEndPuncttrue
\mciteSetBstMidEndSepPunct{\mcitedefaultmidpunct}
{\mcitedefaultendpunct}{\mcitedefaultseppunct}\relax
\EndOfBibitem
\bibitem[Chuang \latin{et~al.}(2011)Chuang, Su, Jeng, Hong, Su, Su, Huang,
  Laio, and Su]{Chuang}
Chuang,~W.-T.; Su,~W.-B.; Jeng,~U.-S.; Hong,~P.-D.; Su,~C.-J.; Su,~C.-H.;
  Huang,~Y.-C.; Laio,~K.-F.; Su,~A.-C. \emph{Macromolecules} \textbf{2011},
  \emph{44}, 1140--1148\relax
\mciteBstWouldAddEndPuncttrue
\mciteSetBstMidEndSepPunct{\mcitedefaultmidpunct}
{\mcitedefaultendpunct}{\mcitedefaultseppunct}\relax
\EndOfBibitem
\bibitem[Imai \latin{et~al.}(1995)Imai, Kaji, Kanaya, and Sakai]{imai1995}
Imai,~M.; Kaji,~K.; Kanaya,~T.; Sakai,~Y. \emph{Phys. Rev. B} \textbf{1995},
  \emph{52}, 12696--12704\relax
\mciteBstWouldAddEndPuncttrue
\mciteSetBstMidEndSepPunct{\mcitedefaultmidpunct}
{\mcitedefaultendpunct}{\mcitedefaultseppunct}\relax
\EndOfBibitem
\bibitem[T.Kawakatsu(2010)]{Kawakatsu}
T.Kawakatsu, \emph{Statistical Physics of Polymers: An Introduction};
  Springer-Verlag, 2010; Chapter 2\relax
\mciteBstWouldAddEndPuncttrue
\mciteSetBstMidEndSepPunct{\mcitedefaultmidpunct}
{\mcitedefaultendpunct}{\mcitedefaultseppunct}\relax
\EndOfBibitem
\bibitem[Leibler(1980)]{Leibler}
Leibler,~L. \emph{Macromolecules} \textbf{1980}, \emph{13}, 1602--1617\relax
\mciteBstWouldAddEndPuncttrue
\mciteSetBstMidEndSepPunct{\mcitedefaultmidpunct}
{\mcitedefaultendpunct}{\mcitedefaultseppunct}\relax
\EndOfBibitem
\bibitem[G.Strobl(2007)]{Strobl}
G.Strobl, \emph{The Physcs of Polymers: Concepts for Undestanding Their
  Structure and Behavior}; Springer-Verlag, 2007; Chapter 2\relax
\mciteBstWouldAddEndPuncttrue
\mciteSetBstMidEndSepPunct{\mcitedefaultmidpunct}
{\mcitedefaultendpunct}{\mcitedefaultseppunct}\relax
\EndOfBibitem
\bibitem[M.Doi and Edwards(1986)M.Doi, and Edwards]{Doi_Edwards}
M.Doi,; Edwards,~S. \emph{The Theory of Polymer Dynamics}; OXFORD SCIENCE
  PUBLICATIONS, 1986; Chapter 2\relax
\mciteBstWouldAddEndPuncttrue
\mciteSetBstMidEndSepPunct{\mcitedefaultmidpunct}
{\mcitedefaultendpunct}{\mcitedefaultseppunct}\relax
\EndOfBibitem
\end{mcitethebibliography}

\end{document}